\documentclass[%
reprint,
superscriptaddress,
 amsmath,amssymb,
 aps,
longbibliography
]{revtex4-1}

\usepackage{graphicx}% Include figure files
\usepackage{bm}% bold math
\usepackage[normalem]{ulem}
\usepackage{xcolor}% Colors
\usepackage{hyperref}% add hypertext capabilities
\definecolor{bleuf}{rgb}{0,0.44,0.72}
\definecolor{bleu}{rgb}{0,0.44,0.72}
\hypersetup{
    colorlinks=true,                          
    linkcolor=bleuf, % Couleur des liens internes
    citecolor=bleuf, % Couleur des numâÃ Ã¶Â¬Â©ros de la biblio dans le corps
    urlcolor=bleuf  } 

\usepackage{placeins} %Float management

\newcommand\grad\nabla
\newcommand\del\nabla
\renewcommand{\b}[1] {\mathbf{#1}}

\newcommand\dd{\mathrm{d}}

\begin{document}

\title{
Emergence of dynamic vortex glasses in disordered polar active fluids
}
\author{Am\'elie Chardac}
\affiliation{Univ. Lyon, ENS de Lyon, Univ. Claude Bernard, CNRS, Laboratoire de Physique, F-69342, Lyon, France}
\author{Suraj Shankar }
\affiliation{Department of Physics, Harvard University, Cambridge, MA 02318, USA}
\author{M. Cristina Marchetti}
\affiliation{Department of Physics, University of California Santa Barbara, Santa Barbara, CA 93106, USA}
\author{Denis Bartolo}
\affiliation{Univ. Lyon, ENS de Lyon, Univ. Claude Bernard, CNRS, Laboratoire de Physique, F-69342, Lyon, France}
\date{\today}

\begin{abstract}
%152 words
In equilibrium, disorder conspires with topological defects to redefine the ordered states of matter in systems as diverse as crystals, superconductors and liquid crystals.
Far from equilibrium, however, the consequences of quenched disorder on active condensed matter remain virtually uncharted. Here, we reveal a state of strongly disordered active matter with no counterparts in equilibrium: a dynamical vortex glass. Combining high-content microfluidic experiments and theory, we show how colloidal flocks collectively cruise through disorder without relaxing the topological singularities of their flows. The resulting state is highly dynamical but the flow patterns, shaped by a finite density of frozen vortices, are stationary and exponentially degenerated. Quenched isotropic disorder acts as a random gauge field turning active liquids into dynamical vortex glasses. We argue that this robust mechanism should shape the collective dynamics of a broad class of disordered active matter, from synthetic active nematics to collections of living cells exploring heterogeneous media.
\end{abstract}

\maketitle

From a physicist perspective, flocks are ensembles of living or synthetic motile units  collectively moving along a common emerging direction~\cite{Toner_Review,Marchetti_review,Cavagna_Review,Chate2020}. 
They realize one of the most robust ordered states of  matter observed over five orders of magnitude in scale and in systems as diverse as motility assays, self-propelled colloids, shaken grains and actual flocks of birds~\cite{Bausch2010,Bricard2013,Yan2016,Granick_Review,Deseigne2012,Sood2014,Cavagna_Review}. 
The quiet flows of flocks are in stark contrast with the spatio-temporal chaos consistently reported and predicted in active nematic liquid crystals, another abundant form of ordered active matter realized in biological tissues, swimming cells, cellular extracts and shaken rods~\cite{Marchetti_review,Doostmohammadi2018}.
Active nematics do not support any form of long range order \cite{shankar2018low,Chate2020}. Their structure is continuously bent and destroyed by the proliferation and annihilation of singularities in their local orientation: topological defects \cite{Doostmohammadi2018,Sanchez2012,giomi2013defect,shankar2018defect}.
Unlike in active nematics, topological defects in flocking matter are merely transient excitations which annihilate rapidly and allow  uniaxial order to extend over system-spanning scales~\cite{Chate2020}. 

This idyllic  view of the ordered phases of active liquids is limited, however, to pure systems. Disorder is known to profoundly alter the stability of topological defects and  the corresponding ordered states in equilibrium  condensed matter~\cite{Blackman1991,deGennes2018,Crabtree1997}, but its role in active fluids remains virtually uncharted territory. 
All previous studies~\cite{Peruani2013,Quint2015,Bechinger2016,Morin2017,Das2018,Reichhardt2018,Toner2018,Toner20182}, including our own early experiments~\cite{Morin2017}, have been limited to weak disorder and smooth perturbations around topologically trivial states. Unlike in equilibrium,  \emph{disorder-induced} topological excitations have been out of reach of any  active-matter experiment, theory and simulations. 

 In this article we show how isotropic disorder generically challenges the extreme robustness of flocking matter to topological defects. We map the full phase behavior of  colloidal flocks in disordered environments and reveal an unanticipated state of  active matter: a dynamical vortex glass.  In dynamical vortex glasses, millions of self-propelled particles can steadily cruise though disorder maintaining local orientational order and  without relaxing the topological singularities of their flows. The associated flow patterns are exponentially degenerated  and shaped by  amorphous ensembles of frozen topological defects, yielding a \emph{dynamical} state akin to the \emph{static} vortex-glass phase of dirty superconductors and random-gauge magnets~\cite{Rubinstein83,Nattermann2000,Carpentier2000}. Building a theory of flock hydrodynamics beyond the spin-wave approximation, we elucidate the emergence and stabilization of topological vortices by quenched disorder. Finally, we discuss the universality of the dynamical vortex glass phase beyond the specifics of polar active matter and colloidal flocks.

\section*{Disordered flocks on a chip}
\begin{figure*} 
	\begin{center}
		\includegraphics[width=17.8cm]{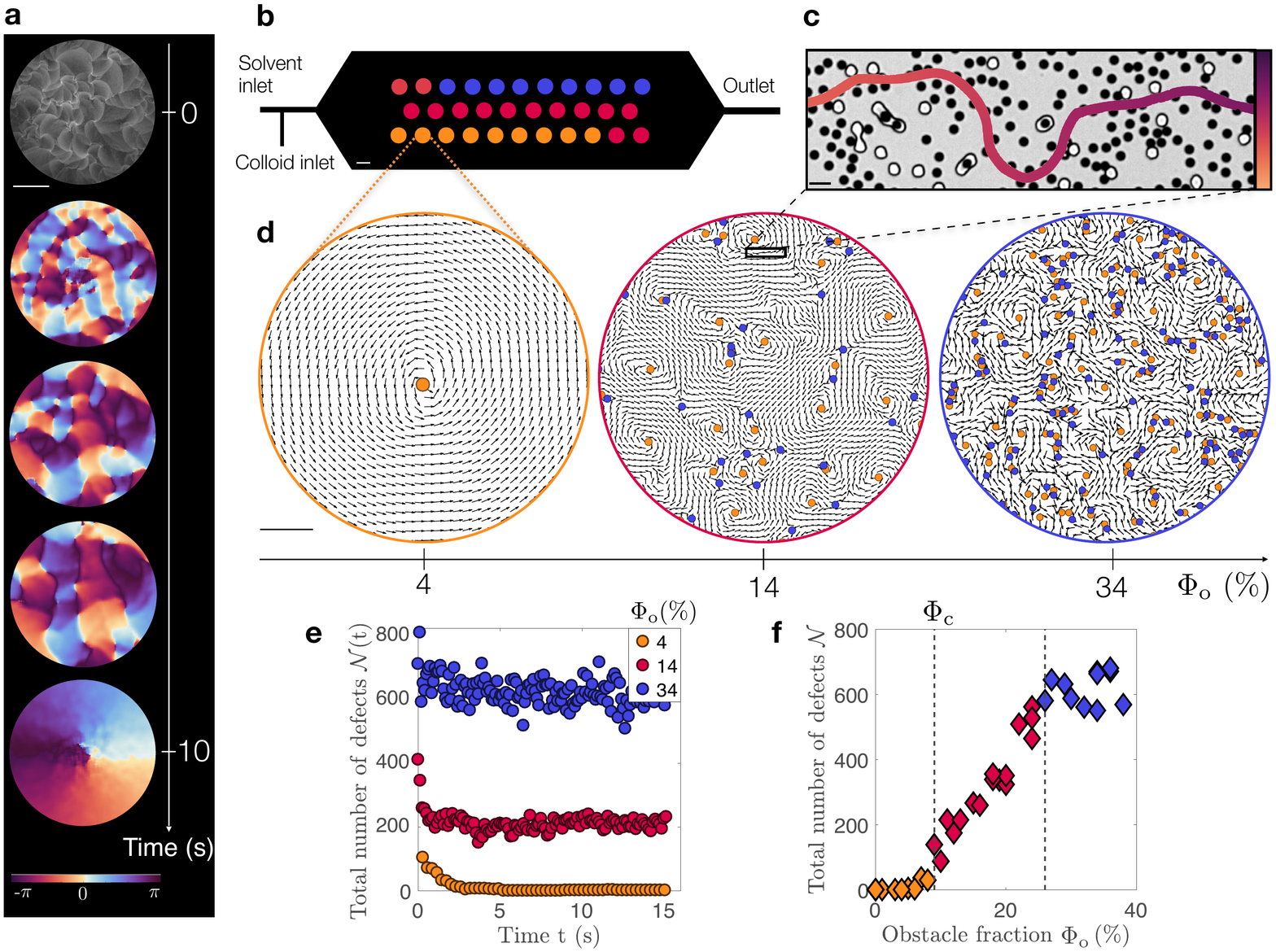} 
		\caption{{\bf Emergence and destruction of polar flows}. 
    {\bf A}, Experimental picture of a Quincke-roller fluid at the onset of self-propulsion and subsequent Schlieren patterns of the flow field. The colormap indicates the angle of the local velocity field with the $x$-axis. The many topological defects formed at the onset of collective motion coarsen yielding a pristine azimuthal flow. Scale bar: $1\,\rm mm$.
    {\bf B}, Schematics of the microfluidic device. 32 circular chambers are patterned with random lattices of circular obstacles. Each chamber correspond to a different value of $\Phi_{\rm o}$. A T-junction makes it possible to continuously vary the packing fraction of Quincke rollers in all chambers at once.
    Same color coding as in {\bf F}. Scale bar: $3\,\rm mm$.
    {\bf C}, Closeup picture showing Quincke rollers collectively moving through a disordered lattice of micro-fabricated obstacles. A roller trajectory shows that the obstacles repel the active colloids at a finite distance. Scale bar: $10\,\rm \mu m$. Color bar: time, from the oldest positions (bright colors) to the latest positions (dark colors). 
   {\bf D}, Polarization fields in the vortex, meander and gas states. The dots indicate the location of the topological defects (Orange: +1, Blue: -1). Scale bar: $0.5\,\rm mm$.
    {\bf E}, Evolution of the number of topological defects ${\mathcal N}(t)$ during the coarsening process for $\Phi_{\rm o}=4\,\%$ (Polar liquid), $\Phi_{\rm o}=14\,\%$ (Meander), $\Phi_{\rm o}=34\,\%$ (Gas). The error in the number of defect measurement is smaller than the symbol size, see SI.    
    {\bf F}, Variations of the average number of topological defects in the steady state with the obstacle fraction. The meander phase emerges at $\Phi_{\rm c}=9\pm1\,\%$.}
		\label{Fig1}
		\end{center}
\end{figure*}

Our experiments are based on the model system we introduced in~\cite{Bricard2013}. We use Quincke rotation to power inanimate polystyrene beads of radius $a=2.4\,\rm \mu m$~\cite{Quincke,Lavrentovich2016}, and turn them into colloidal rollers self-propelling at constant speed $\sim1\,\rm mm/s$ in circular microfluidic chambers of radius $R=1.5\,\rm mm$, see Fig.~\ref{Fig1}A and Movie S1. Before addressing the impact of disorder on their collective dynamics, it is worth recalling the phenomenology of the pure ordered phase~\cite{Bricard2015,Geyer2019}.

As the rollers are set in motion, they interact and self-assemble into a spontaneously flowing liquid equivalently referred to as a flock, a Toner--Tu fluid, or a polar liquid~\cite{Toner95,Marchetti_review} shown in Movie S1.  
The resulting flow patterns are initially isotropic and marred by a number of $\pm1$ topological defects clearly visible in the Schlieren patterns of Fig.~\ref{Fig1}A and Movie S2. However, the velocity-alignment interactions responsible for flocking motion penalize flow distortions thereby causing the attraction and annihilation of defects of opposite charge. The resulting lively coarsening dynamics lasts few tens of seconds and ultimately yields  pristine azimuthal flows \emph{with long-range order in two dimensions}, Fig.~\ref{Fig1}A. 
%%%%

%%%%
%%%

To comprehensively investigate how disorder alters the phases of polar active matter, we perform high-content experiments on the microfluidic device sketched in Fig.~\ref{Fig1}B. On a single chip, we  quantify the flows of disordered roller fluids in 32 different disordered geometries. Disorder is implemented in 3 mm wide circular chambers decorated by random arrays of isotropic obstacles of diameter $10\,\rm \mu m$ that repel the rollers at a distance, while leaving their speed unaltered, see Fig.~\ref{Fig1}C and Movie S3. We replicate the experiments increasing the obstacle fraction $\Phi_{\rm o}$ from $0\,\%$ to $38\,\%$ keeping the system below the Lorentz localization transition of individual rollers~\cite{Zeitz2017,Morin2017b}. All experiments are systematically performed for several disorder realizations and initial conditions, see also Methods.

Let us first discuss the impact of disorder on a polar liquid  deep in the flocking phase, for a colloid number density $\rho_0=7\pm0.5\times10^{-2}$. For small disorder, $\Phi_{\rm o}<9\,\%$, the  obstacles hardly disrupt the collective flow, as $\pm 1$ defects formed at the onset of collective motion quickly annihilate. The system coarsens to an ordered flocking state which, in the disk geometry, takes the form of a macroscopic vortex centered around a single $+1$ defect, see Fig.~\ref{Fig1} and Movie S4.
In stark contrast, increasing $\Phi_{\rm o}$ above $9\,\%$, a very distinct type of organization emerges. Disorder essentially arrests the coarsening by pinning $\pm 1$ defects. While the global flow vanishes, we find locally correlated flows through meandering patterns bent by disorder (Fig.~\ref{Fig1}D and Movies S5 and S6). Particles stream coherently through the meanders, resulting in a finite correlation length for orientational order, but the structure of the flow network itself is static and fixed in space, as shown in Fig.~\ref{Fig1}D and Movie S6.
 Further increasing the obstacle fraction (above $26\,\%$), we recover the liquid-gas transition reported in~\cite{Morin2017}. In the gas phase, particles are motile, but there is no orientational order as
 defect pairs continuously unbind and annihilate as seen in Figs.~\ref{Fig1}D,~\ref{Fig1}E. % and Movie S7.
  The two transitions between the three dynamical states are unambiguously determined by the variations of $\mathcal N(\Phi_{\rm o})$, the mean number of topological defects in the velocity field, plotted in Fig.~\ref{Fig1}F (the detection of the topological defects is detailed in Supplementary Note 3). The emergence of meandering flows is signaled by the linear increase of $\mathcal N$ as $\Phi_{\rm o}$ exceeds the critical value $\Phi_{\rm c}=9\,\%$, whereas the loss of local orientational order saturates $\mathcal N(\Phi_{\rm o})$ above $\Phi_{\rm o}=26\,\%$.\\ 

\section*{Phase behaviour of disordered polar active matter}

\begin{figure*}
	\begin{center}
    \includegraphics[width=17.8cm]{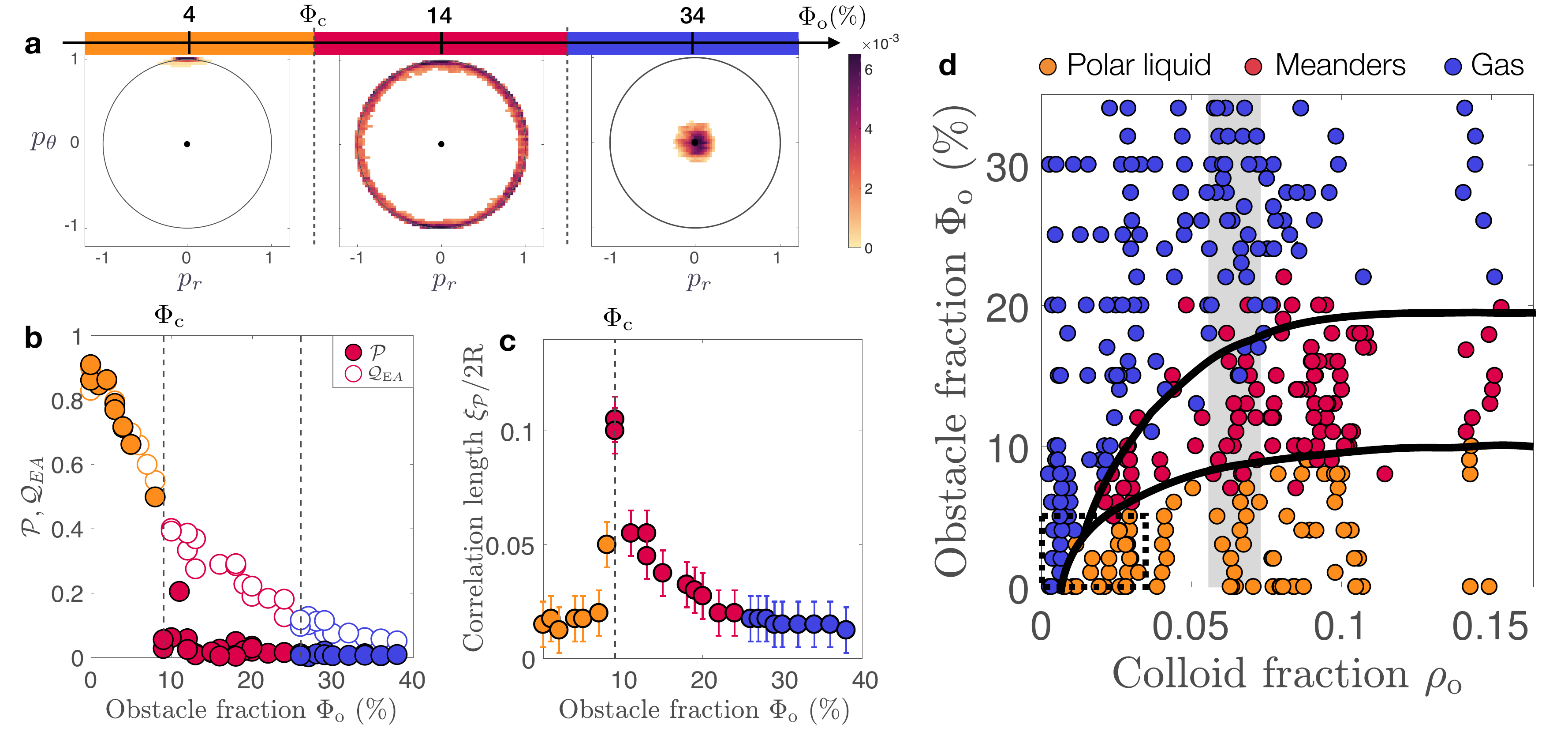}
 \caption{{\bf Suppression of long-range orientational order}.  
{\bf A}, PDF of the polarization fields $\mathrm{P}(p_r,p_{\theta})$ in the three dynamical phases. Note that the distribution is isotropic but peaked along the unit circle in the meander phase: spontaneous flows are locally preserved but global polar order is suppressed by disorder.
{\bf B}, Variation of the global polarization $\mathcal P$ and Edwards--Anderson $\mathcal Q_{\rm EA}$ order parameters. In  polar liquids $\mathcal P>0$, in meanders $\mathcal P=0$ but $\mathcal Q_{\rm EA}>0$, and in a gas both $\mathcal P$ and $\mathcal Q_{\rm EA}$ vanish. 
{\bf C}, Correlation length of the polarization field defined from the exponential decay of $\mathcal P(\ell)$  plotted in Supplementary Figure S1. The cusp of the $\xi_{\mathcal P}(\Phi_{\rm o})$ curve at $\Phi_{\rm c}$ hints towards a critical dynamical transition. The accuracy in the measurement of $\xi/2R$ is of the order of $10^{-2}$.
 {\bf D}, Phase diagram of polar active matter in isotropic disorder. The phase boundaries are equivalently determined from the number of topological defects (Fig.~\ref{Fig1}F), or from the variations of $\mathcal P$ and $\mathcal Q_{\rm EA}$. In practice, we define the boundary separating the polar-liquid from the meander phase as $\mathcal P>0.45$ and the boundary between the meander and gas phases as $\mathcal P<0.45$ and $Q_{\rm EA}>0.2$. The geometry of the phase boundaries does not crucially depend on these criteria. The extent of the meander phase becomes vanishingly small only at the onset of the flocking transition. The dashed rectangle indicates the region of the phase diagram explored in~\cite{Morin2017} where the meander phase was missed. Orientational order is generically lost due to the emergence of meander patterns. The gray rectangle corresponds to the series of experiments discussed in the main text.  
} 
		\label{Fig2}
		\end{center}
\end{figure*}

In order to quantitatively distinguish the dynamics in the three regimes, we inspect the statistics of the time-averaged polarization defined as $\bm p(\mathbf r)=\langle \hat {\mathbf v}(\mathbf r,t)\rangle_t$, where the unit vector $\hat {\mathbf v}(\mathbf r,t)$ is the instantaneous orientation of the velocity field, see Methods. The distributions of $\bm p(\mathbf r)$ in Fig.~\ref{Fig2}A show that, for small disorder, the flow is uniformly polarized along the azimuthal direction, which reflects nearly perfect polar order.
Conversely, above $\Phi_{\rm c}$ global polar order is suppressed, and the polarization is isotropically distributed on the scale of the system. The strong localization of the distribution on the unit circle demonstrates, however, that meanders persistently distort the streamlines \emph{without} arresting the local flows.
Above $\Phi_{\rm o}=26\,\%$, in the active-gas phase, the typical polarization vanishes, polar order melts and flows are suppressed at all scales.

We now demonstrate that the two flowing patterns correspond to two distinct dynamical phases. We first stress that polar liquids with genuine long range orientational order survive at finite disorder (see also Supplementary Note 6). Unlike active nematics, deconfining the system does not suppress the stationary vortex patterns shown in Fig.~\ref{Fig1}D~\cite{Wioland2016,Wu2017,Opathalage2019}. This crucial result follows from the finite-size analysis of the polarization order parameter $\mathcal P(\ell) =\sqrt{\langle p_r (\mathbf r)\rangle_{\mathbf r}^2+\langle p_\theta (\mathbf r) \rangle_{\mathbf r}^2}$, where $\ell$ is the size of the region where spatial averaging is performed. This definition of $\mathcal P$ is natural in a circular geometry where the spatial average of $\bm p$ vanishes, even in pure systems. Below $\Phi_{\rm c}$, Supplementary Fig. S1 clearly indicates that $\mathcal P(\ell)$ converges to a finite value over a finite length scale $\xi_{\mathcal P}$, signaling long range polar order. Conversely, deep in the meandering regime $\mathcal P(\ell)$ vanishes over a finite scale. Polar liquids and meanders are therefore two genuinely distinct phases of active matter. 

The transition between these two dynamical phases is captured by the global polarization $\mathcal P\equiv\mathcal P(\ell=2R)$, where $R$ is the chamber radius. We can see in Fig.~\ref{Fig2}B that $\mathcal P$ decays weakly and linearly with $\Phi_{\rm o}$ for small disorder, but drops sharply to $0$ at the critical value $\Phi_{\rm c}$ defined from the proliferation of quenched topological defects in Fig~\ref{Fig1}F. The bifurcation of $\mathcal P$ in Fig.~\ref{Fig2}B suggests a critical scenario. This hypothesis is further supported by Fig.~\ref{Fig2}C which reveals a divergence, or at the very least a drastic increase of the correlation length $\xi_{\mathcal P}$ at the onset of meandering motion.
The polarization $\mathcal P$ alone, however, does not distinguish the meander from the gas phase. 
We thus introduce the Edwards--Anderson parameter $\mathcal{Q_{\rm EA}}=\langle \hat{\mathbf v}(\mathbf r,t) \cdot \hat{\mathbf v}(\mathbf r,t+T) \rangle_{\mathbf r,t,T\to\infty}$ that quantifies the temporal persistence of the emergent flows~\cite{Cavagna2009}. The finite non-zero value of $\mathcal{Q_{\rm EA}}$ for $\Phi_{\rm c}<\Phi_{\rm o}<26\,\%$ (Fig.~\ref{Fig2}B) confirms the persistence of polar order along the meanders. We also find that the continuous transition, or crossover, between the meander and the gas phase, where $\mathcal{Q_{\rm EA}}\sim0$, coincides with that identified from the topological defect statistics in Fig.~\ref{Fig1}F.
%

%%%

Taking advantage of our high-content microfluidic experiments, we measure $\mathcal P$ and $\mathcal Q_{\rm EA}$ in more than 400 experiments performed varying the roller fraction from $10^{-3}$ to $1.5\times10^{-1}$, and $\Phi_{\rm o}$ from 0 to $38\,\%$ for multiple disorder realizations and initial conditions. The resulting phase diagram shown in Fig.~\ref{Fig2}D firmly establishes that disorder and topological defects do not merely offset the flocking transition as reported in~\cite{Morin2017}, but conspire to bend the dynamics of polar liquids  into amorphous and singular flows while preserving local flocking motion.\\

\section*{Meander flows are dynamical vortex glasses}

\begin{figure*} 
	\begin{center}
	\includegraphics[width=17.8cm]{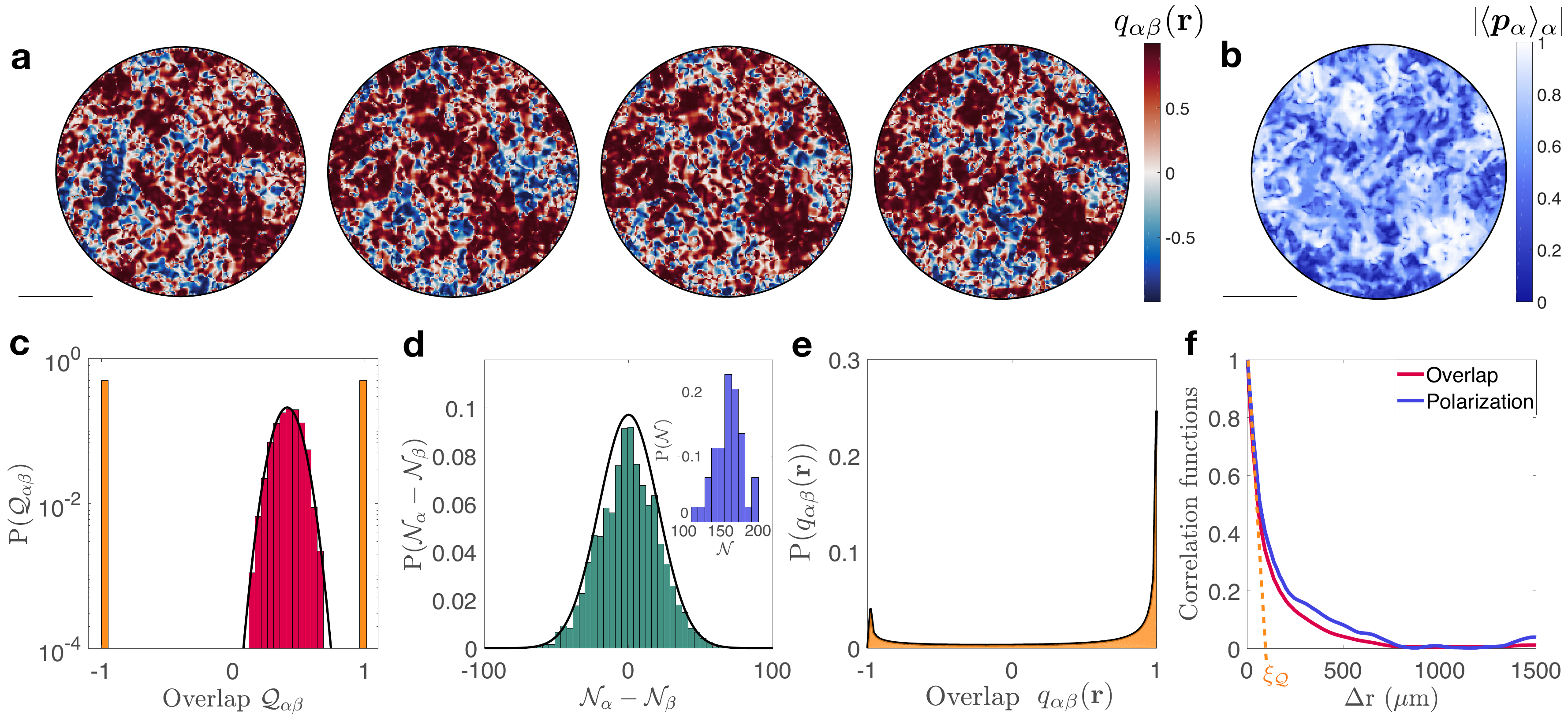}
		\caption{{\bf Meanders as dynamical vortex glasses.}
		{\bf A}, Maps of the overlap field $q_{\alpha \beta}(\mathbf{r})$ for 4 out of 1225 different pairs of replicas ($\alpha$, $\beta$). Scalebar: $1\,\rm mm$.
		{\bf B}, Magnitude of the replica-averaged polarization field. Disorder robustly sets the flow orientation in macroscopic regions (light colors). Scalebar: $1\,\rm mm$.
		{\bf C}, Overlap distributions in the polar liquid and meander phases. In the polar liquid phase (light orange) the distribution is symmetric and bimodal. Clockwise and counterclockwise vortices are equiprobable. In the meander phase, the distribution is a wide Gaussian centered on $0.46$. 
		{\bf D}, Distribution of the difference between the number of topological defects in each pair of replicas. The distribution is a Gaussian with a standard deviation $\sigma=18\pm3$ much larger than the uncertainty defined  by the standard deviation of the topological charge (which should be constrained to equal 1 in a circular geometry). Inset: distribution of the number of defects in each replica. The replicated flows are topologically distinct.  
		{\bf E}, Probability density of the local overlap $q_{\alpha\beta}(\mathbf{r})$. The distribution is asymmetric but sharply peaked at $\pm1$. Locally the active fluid either flows along identical or opposite directions in each replica of the same experiment.
		{\bf F}, Decay of the spatial correlation of $q_{\alpha\beta}(\mathbf{r})$ averaged over all replicas and of the replica-averaged polarization field plotted in ({\bf B}). Both decorrelations occur over the same finite distance $\xi_{\mathcal Q}=100\pm10\,\rm \mu m$. All panels correspond to experiments performed deep in the meander phase $\Phi_{\rm o}=15\,\%$.
} 
		\label{Fig3}
		\end{center}
\end{figure*}

We now establish that the meanders realize a dynamical vortex glass.
To do so, we compare the flow orientations of fifty replicas of the very same experiment ($\Phi_{\rm o}=15\,\%$). We solely vary the initial conditions, keeping the colloid fraction and disorder realization identical.
 We then quantify the  resemblance between the flow patterns by introducing a measure of the local overlap $q_{\alpha \beta}(\mathbf{r})= {\bm p}_{\alpha}(\mathbf{r}) \cdot {\bm p}_{\beta}(\mathbf{r})$ between replicas $\alpha$ and $\beta$. Examples of overlap maps are shown in Fig.~\ref{Fig3}A for four different pairs of experiments, see also Supplementary Fig.~S5. 
 The replicated flows are identical when $q_{\alpha \beta}(\mathbf{r})=+1$, opposite when $q_{\alpha \beta}(\mathbf{r})=-1$ and orthogonal when $q_{\alpha \beta}(\mathbf{r})=0$. 
A simple inspection of the maps readily indicates that, although all experiments correspond to overall different patterns, a few macroscopic regions are virtually identical from one replica to another. 
More quantitatively, we plot in Fig.~\ref{Fig3}C the distribution of the global overlap ${\mathcal Q}_{\alpha\beta}=\langle q_{\alpha\beta}(\mathbf r)\rangle_{\mathbf r}$, and compare it to that of the macroscopic vortex obtained in the flocking state. 
The two distributions are markedly different. In the flocking state, $\rm P({\mathcal Q}_{\alpha\beta})$ is composed of two symmetric peaks at ${\mathcal Q}_{\alpha\beta}=\pm1$, as global circulation of either handedness are equally probable. Conversely, in the meander phase, $\rm P({\mathcal Q}_{\alpha\beta})$ is not symmetric. It is a broad Gaussian distribution centered around a positive mean value. In other words, even though disorder is isotropic and homogeneous, the obstacles bias the orientation of the flow field over macroscopic regions of space, as a result reversing the orientation of the flow does not yield an equally probable pattern. The bias in the flow structure is clearly visible in the replica-averaged velocity field shown in Fig~\ref{Fig3}B. 

Crucially, the large width of the overlap distribution demonstrates a broad range of meandering patterns that do not merely differ from each other by continuous distortions. The replicas are actually \emph{topologically} inequivalent, as demonstrated by the distribution of the difference in the number of topological-defect between each pair of replicas (${\mathcal N}_\alpha-{\mathcal N}_\beta$). The regions of the flow patterns that are robust to variations in the initial conditions are therefore shaped by pinned defects of identical charge and orientation which persist in all replicas. In other words, defects are strongly correlated not only in their spatial location, but also in their flow direction. This becomes evident upon inspecting the  statistics of local overlap.

Unlike the global overlap distribution, $\mathrm{ P}(q_{\alpha\beta}(\mathbf r))$ is bimodal and sharply peaks at $q_{\alpha\beta}(\mathbf r)=\pm1$, see Fig.~\ref{Fig3}E. 
Simply put, the meanders in each pair of replicas differ by the reversal of the flow orientation over a finite fraction of space. This area fraction is given by $1 - \left[\mathrm{ P(}q_{\alpha\beta}=+1)-\mathrm{P(}q_{\alpha\beta}=-1)\right]\sim 0.5$, see also Supplementary Note III. We can then estimate the typical extent of the compact regions where the flow can flip sign from one replica to another by measuring  the correlation length of the local overlap, or of the replica-averaged flow shown in Fig.~\ref{Fig3}F. We find that the overlap decorrelates exponentially over a finite distance $\xi_{\mathcal Q}=100\pm10\,\rm\mu m$. This finite correlation length implies that the meanders explore a conformational landscape including a number of steady states that scales as  $\sim 2^{(R/\xi_{\mathcal Q})^2}$. The exponential degeneracy of the flows demonstrates that the meander phase is the dynamical analogue of a topological defect glass. It is strongly reminiscent of the vortex glass phases found in disordered flux lines in superconductors and in random-gauge Heisenberg magnets~\cite{Fisher89,Nattermann2000,Carpentier2000}.\\

\section*{Dynamical vortex glass as a gauge glass}

We now make a precise connection between the dynamical vortex glass we observe here in active fluids and its static counterpart in superconductors.

To do so, we construct a hydrodynamic theory of disordered polar flows, building on a final set of experiments performed in periodic lattices of obstacles, see Supplementary Fig. S7 and Movie S7. A systematic investigation reveals that no meander forms in the absence of disorder, see Supplementary Note 5. The spatial heterogeneities in the local obstacle fraction $\phi_{\rm o}(\mathbf r)$ are essential to generate the meandering flow patterns. At lowest order in gradients, disorder thus minimally alters Toner--Tu hydrodynamics for the conserved roller density $\rho$ and the flow velocity $\mathbf{v}$~\cite{Toner95,Toner_Review} ,
\begin{gather}
\partial_t\rho=-\bm\nabla\cdot(\rho\mathbf{v}), \;
\label{Re.rho}\\
\partial_t \mathbf v+\lambda \mathbf v\cdot \bm \nabla\mathbf v=(a_2-a_4|\mathbf{v}|^2)\mathbf v+K\nabla^2 \mathbf v-\beta \bm \nabla \rho -\beta_{\rm o}\bm \nabla \phi_{\rm o}\ ,
\label{Eq.TT}
\end{gather}

All the transport coefficients are taken to be constant positive parameters.
 When $\bm \nabla \phi_{\rm o}=\bm 0$, Eq.~\ref{Eq.TT} reduces to the standard Toner--Tu equations, which quantitatively predicts the vortical flows of Fig.~\ref{Fig1}A~\cite{Bricard2015}. {The  terms $a_2,\,a_4$ set the mean value of the flow velocity ($v_0=\sqrt{a_2/a_4}$) deep in the homogeneous ordered phase, while $\lambda$ controls self-advection and the elastic constant $K$ penalizes flow distortions.} The $\beta_{\rm o}$ term  reflects the coupling to disorder, and provides a direct qualitative explanation for the bias of the overlap distribution in Fig.~\ref{Fig3}C. Although the obstacles are isotropic and, on average, homogeneously distributed, $\phi_{\rm o}(\mathbf r)$ acts as a random pressure field that locally drives the flows towards the same obstacle-depleted regions in all replicas. {For weak disorder ($\Phi_{\rm o}<\Phi_c$), the ordered flows are only marginally perturbed by the obstacles. As detailed in Supplementary Note 6, we can then  perform a simple linearized analysis about the polarized state, deep in the order phase where $\mathbf{v}=v_0(\cos\theta,\sin\theta)$. We  readily obtain the steady state orientational correlator in Fourier space
\begin{gather}
    \overline{\theta_{\b{q}}^2}=\dfrac{\beta_{\rm o}^2q_{x}^2q_{y}^2}{[(c_{\parallel}v_0q_{x}^2-c_{\perp}^2q_{y}^2)^2+v_0^2K^2q_{x}^2q^4]}\Phi_{\rm o}\ ,
\end{gather}
where the overbar denotes a disorder average.  $c_{\parallel}=\lambda v_0$ and $c_{\perp}=\sqrt{\beta\rho_0}$ are the longitudinal and transverse sound speeds respectively, and $\rho_0$ is the mean colloidal roller density. Velocity fluctuations hence remain finite on large scales signaling true long-range order. The mean polarization merely decays linearly with $\Phi_{\rm o}$, in agreement with our experimental measurements in the small disorder limit ($\Phi_{\rm o}<\Phi_{\rm c}$), see Fig.~\ref{Fig2}C and Supplementary Note 6.

For strong disorder, we have to go beyond this spin-wave theory and account for both nonlinearities and the presence of topological defects. Unlike in equilibrium  ferromagnets  and active-nematic films, 2D colloidal flocks support genuine long range order, thereby ruling out the classical Kosterlitz-Thouless picture of defect unbinding. Topological defects in clean polar flocks only participate in a rapid coarsening process by annihilating quickly to yield pristine ordered flow, as evidenced in Fig.~\ref{Fig1}A.   
Disorder changes this picture in important ways. Quenched fluctuations due to the random obstacles provides an additional contribution that competes with orientational elasticity to uncover a new active defect-unbinding scenario. For simplicity, we fix the colloidal roller density to be a constant ($\rho=\rho_0$); the sole hydrodynamic mode, is therefore the orientation $\theta$. As detailed in Supplementary Notes 7 and 8, retaining the full nonlinear dependence on $\theta$ in Toner-Tu equations, we find that at long-wavelengths the convective nonlinearities solely compete with the random pressure term in Eq.~\ref{Eq.TT}. To gain some insight on this competition, let us first consider a local minimum of $\phi_{\rm o}(\mathbf r)$ at the origin, the inward disorder pressure must be balanced by a centrifugal kinematic pressure $\sim\lambda v^2/r$ associated with an azimuthal flow. In a more general setting, the competition between convection and random pressure constrain the angular fluctuations to obey,
\begin{equation}
    \bm\del\theta=-\dfrac{\beta_{\rm o}}{\lambda v_0^2}\hat {\b{z}}\times \mathbb{P}\cdot\bm\del\phi_{\rm o}\equiv\mathcal{A}(\b{r})\ ,\label{eq:gauge}
\end{equation}
where $\mathbb{P}=\b{1}-\hat{\b{v}}\hat{\b{v}}$. Remarkably, this quenched distribution of phase distortions is reminiscent of random-gauge $XY$ models, and flux-line glasses in type II superconductors, see e.g.~\cite{Rubinstein83,Nattermann2000,Carpentier2000}. Eq.~\eqref{eq:gauge} reveals that active self-advection nonlinearly transforms the quenched disorder potential $\phi_{\rm o}$ into an effective random gauge vector field $\mathcal{A}$.  The  local circulation of $\mathcal{A}$ then defines a \emph{quenched} topological charge in the background $Q(\b{r})=\hat{\b{z}}\cdot\left[\bm\del\times\mathcal{A}(\b{r})\right]$.

 For a given disorder configuration, the flow streamlines on large scales must satisfy the nonlinear constraint in Eq.~\ref{eq:gauge}, which can be used to obtain  the steady state distribution of $\theta$ to be $\mathrm{P}[\theta(\b{r})]\propto\lim_{T\to 0}\exp(-E/T)$ with an auxiliary temperature $T$ and an effective energy (see Supplementary Note 7 for details)
\begin{equation}
    E[\theta(\b{r})]=\dfrac{K}{2}\int\dd^2r\;|\bm\del\theta-\mathcal{A}(\b{r})|^2\ ,\label{eq:E}.
\end{equation}
The zero temperature ground state controls the flow configurations of the disordered flock  ($T\to 0$). This completes our mapping of the nonlinear meandering flows in a polar fluid on a disordered substrate to the well known problem of finding the ground states of a random-gauge $XY$ model \cite{Rubinstein83,carpentier1998disordered,Carpentier2000}.

The effective energy in Eq.~\ref{eq:E} captures the competition between disorder generating meanders dictated by $\mathcal{A}$ and elastic restoring forces penalizing these distortions. This allows a Kosterlitz-Thouless style argument for proliferating free vortices detailed in Supplementary Note 8.  We stress, however, on the crucial role of activity in our analysis. Although we obtain an equilibrium-like mapping in Eq.~\ref{eq:E}, activity is crucial in generating the gauge field $\mathcal{A}$ via self-advection.  Noting $a$ the vortex-core size, the elastic penalty of an isolated defect $\sim K\ln(R/a)$  can be offset by the energy gain $\sim K\Phi_{\rm o}(\beta_{\rm o}/\lambda v_0^2)\ln(R/a)$ from optimally screening out the background charge. Their balance hence predicts a critical threshold $\Phi_{\rm c}\propto\lambda v_0^2/\beta_{\rm o}$ beyond which pinned vortices proliferate, in agreement with our measurements of the number of  frozen defects at the onset of of the vortex glass formation, Fig.~\ref{Fig1}F.

 A more sophisticated analysis including renormalization corrections to the elastic constant is presented in Supplementary Note 8. This rationalizes the existence of the dynamical vortex glass phase along with its phase transition as being driven by  disorder driven unbinding of $\pm1$ vortices. 

\section*{Concluding remarks}
In conclusion, combining high-content experiments and theory we have shown how dynamical vortex glasses generically emerge when the polar order of flocking matter competes with isotropic disorder. 
We emphasize the generality of our predictions. 
Both the topological-defect stabilisation scenario and the extensive complexity of the flow patterns solely rely on stable uniaxial order in a fluid assembled from motile units. Beyond the specifics, of collodial active matter, we therefore expect dynamical vortex glasses to emerge in a host of active materials ranging from confined microtubule nematics~\cite{Doostmohammadi2018}, to concentrated bacteria suspensions~\cite{Wioland2016}, and cell tissues~\cite{Duclos2018} cruising through disorder. Establishing a quantitative theory of their amorphous patterns remains a formidable challenge.\\

\noindent {\bf Acknowledgements.} We acknowledge support from ANR program WTF and Idex program ToRe. M.~C.~M.~was supported by the US National Science Foundation through award DMR-1938187. S.~S.~is supported by the Harvard Society of Fellows. We thank D.~Carpentier, O.~Dauchot, W. T. M. Irvine, and A.~Morin  for insightful comments and suggestions.

\noindent {\bf Data availability} The data that support the plots within this paper and other findings of this study are available from the corresponding author upon request.
 
\noindent{\bf Author Contributions} D.~B. conceived the project. A.~C. and D.~B. designed the experiments. A.~C. performed the experiments and analyzed data. A.~C. and D.~B. interpreted and discussed the experiments. M.~C.~M. and S.~S. performed the theory. All authors discussed the results and wrote the manuscript. A.~C. and S.~S. contributed equally.

\noindent{\bf Author Information} Correspondence and requests for materials
should be addressed to D.~B. (email: denis.bartolo@ens-lyon.fr).  

\noindent{\bf Competing interests} The authors declare no competing interests.

\bibliography{Biblio}

\begin{thebibliography}{48}%
\makeatletter
\providecommand \@ifxundefined [1]{%
 \@ifx{#1\undefined}
}%
\providecommand \@ifnum [1]{%
 \ifnum #1\expandafter \@firstoftwo
 \else \expandafter \@secondoftwo
 \fi
}%
\providecommand \@ifx [1]{%
 \ifx #1\expandafter \@firstoftwo
 \else \expandafter \@secondoftwo
 \fi
}%
\providecommand \natexlab [1]{#1}%
\providecommand \enquote  [1]{``#1''}%
\providecommand \bibnamefont  [1]{#1}%
\providecommand \bibfnamefont [1]{#1}%
\providecommand \citenamefont [1]{#1}%
\providecommand \href@noop [0]{\@secondoftwo}%
\providecommand \href [0]{\begingroup \@sanitize@url \@href}%
\providecommand \@href[1]{\@@startlink{#1}\@@href}%
\providecommand \@@href[1]{\endgroup#1\@@endlink}%
\providecommand \@sanitize@url [0]{\catcode `\\12\catcode `\$12\catcode
  `\&12\catcode `\#12\catcode `\^12\catcode `\_12\catcode `\%12\relax}%
\providecommand \@@startlink[1]{}%
\providecommand \@@endlink[0]{}%
\providecommand \url  [0]{\begingroup\@sanitize@url \@url }%
\providecommand \@url [1]{\endgroup\@href {#1}{\urlprefix }}%
\providecommand \urlprefix  [0]{URL }%
\providecommand \Eprint [0]{\href }%
\providecommand \doibase [0]{http://dx.doi.org/}%
\providecommand \selectlanguage [0]{\@gobble}%
\providecommand \bibinfo  [0]{\@secondoftwo}%
\providecommand \bibfield  [0]{\@secondoftwo}%
\providecommand \translation [1]{[#1]}%
\providecommand \BibitemOpen [0]{}%
\providecommand \bibitemStop [0]{}%
\providecommand \bibitemNoStop [0]{.\EOS\space}%
\providecommand \EOS [0]{\spacefactor3000\relax}%
\providecommand \BibitemShut  [1]{\csname bibitem#1\endcsname}%
\let\auto@bib@innerbib\@empty
%</preamble>
\bibitem [{\citenamefont {Toner}\ \emph {et~al.}(2005)\citenamefont {Toner},
  \citenamefont {Tu},\ and\ \citenamefont {Ramaswamy}}]{Toner_Review}%
  \BibitemOpen
  \bibfield  {author} {\bibinfo {author} {\bibfnamefont {John}\ \bibnamefont
  {Toner}}, \bibinfo {author} {\bibfnamefont {Yuhai}\ \bibnamefont {Tu}}, \
  and\ \bibinfo {author} {\bibfnamefont {Sriram}\ \bibnamefont {Ramaswamy}},\
  }\bibfield  {title} {\enquote {\bibinfo {title} {Hydrodynamics and phases of
  flocks},}\ }\href {\doibase https://doi.org/10.1016/j.aop.2005.04.011}
  {\bibfield  {journal} {\bibinfo  {journal} {Annals of Physics}\ }\textbf
  {\bibinfo {volume} {318}},\ \bibinfo {pages} {170 -- 244} (\bibinfo {year}
  {2005})},\ \bibinfo {note} {special Issue}\BibitemShut {NoStop}%
\bibitem [{\citenamefont {Marchetti}\ \emph {et~al.}(2013)\citenamefont
  {Marchetti}, \citenamefont {Joanny}, \citenamefont {Ramaswamy}, \citenamefont
  {Liverpool}, \citenamefont {Prost}, \citenamefont {Rao},\ and\ \citenamefont
  {Simha}}]{Marchetti_review}%
  \BibitemOpen
  \bibfield  {author} {\bibinfo {author} {\bibfnamefont {M.~C.}\ \bibnamefont
  {Marchetti}}, \bibinfo {author} {\bibfnamefont {J.~F.}\ \bibnamefont
  {Joanny}}, \bibinfo {author} {\bibfnamefont {S.}~\bibnamefont {Ramaswamy}},
  \bibinfo {author} {\bibfnamefont {T.~B.}\ \bibnamefont {Liverpool}}, \bibinfo
  {author} {\bibfnamefont {J.}~\bibnamefont {Prost}}, \bibinfo {author}
  {\bibfnamefont {Madan}\ \bibnamefont {Rao}}, \ and\ \bibinfo {author}
  {\bibfnamefont {R.~Aditi}\ \bibnamefont {Simha}},\ }\bibfield  {title}
  {\enquote {\bibinfo {title} {Hydrodynamics of soft active matter},}\ }\href
  {\doibase 10.1103/RevModPhys.85.1143} {\bibfield  {journal} {\bibinfo
  {journal} {Rev. Mod. Phys.}\ }\textbf {\bibinfo {volume} {85}},\ \bibinfo
  {pages} {1143--1189} (\bibinfo {year} {2013})}\BibitemShut {NoStop}%
\bibitem [{\citenamefont {Cavagna}\ and\ \citenamefont
  {Giardina}(2014)}]{Cavagna_Review}%
  \BibitemOpen
  \bibfield  {author} {\bibinfo {author} {\bibfnamefont {Andrea}\ \bibnamefont
  {Cavagna}}\ and\ \bibinfo {author} {\bibfnamefont {Irene}\ \bibnamefont
  {Giardina}},\ }\bibfield  {title} {\enquote {\bibinfo {title} {Bird flocks as
  condensed matter},}\ }\href
  {https://doi.org/10.1146/annurev-conmatphys-031113-133834} {\bibfield
  {journal} {\bibinfo  {journal} {Annu. Rev. Condens. Matter Phys.}\ }\textbf
  {\bibinfo {volume} {5}},\ \bibinfo {pages} {183--207} (\bibinfo {year}
  {2014})}\BibitemShut {NoStop}%
\bibitem [{\citenamefont {Chat{\'e}}(2020)}]{Chate2020}%
  \BibitemOpen
  \bibfield  {author} {\bibinfo {author} {\bibfnamefont {Hugues}\ \bibnamefont
  {Chat{\'e}}},\ }\bibfield  {title} {\enquote {\bibinfo {title} {Dry aligning
  dilute active matter},}\ }\href@noop {} {\bibfield  {journal} {\bibinfo
  {journal} {Annual Review of Condensed Matter Physics}\ }\textbf {\bibinfo
  {volume} {11}},\ \bibinfo {pages} {189--212} (\bibinfo {year}
  {2020})}\BibitemShut {NoStop}%
\bibitem [{\citenamefont {Schaller}\ \emph {et~al.}(2010)\citenamefont
  {Schaller}, \citenamefont {Weber}, \citenamefont {Semmrich}, \citenamefont
  {Frey},\ and\ \citenamefont {Bausch}}]{Bausch2010}%
  \BibitemOpen
  \bibfield  {author} {\bibinfo {author} {\bibfnamefont {Volker}\ \bibnamefont
  {Schaller}}, \bibinfo {author} {\bibfnamefont {Christoph}\ \bibnamefont
  {Weber}}, \bibinfo {author} {\bibfnamefont {Christine}\ \bibnamefont
  {Semmrich}}, \bibinfo {author} {\bibfnamefont {Erwin}\ \bibnamefont {Frey}},
  \ and\ \bibinfo {author} {\bibfnamefont {Andreas~R}\ \bibnamefont {Bausch}},\
  }\bibfield  {title} {\enquote {\bibinfo {title} {Polar patterns of driven
  filaments},}\ }\href
  {http://www.nature.com/nature/journal/v467/n7311/abs/nature09312.html}
  {\bibfield  {journal} {\bibinfo  {journal} {Nature}\ }\textbf {\bibinfo
  {volume} {467}},\ \bibinfo {pages} {73} (\bibinfo {year} {2010})}\BibitemShut
  {NoStop}%
\bibitem [{\citenamefont {Bricard}\ \emph {et~al.}(2013)\citenamefont
  {Bricard}, \citenamefont {Caussin}, \citenamefont {Desreumaux}, \citenamefont
  {Dauchot},\ and\ \citenamefont {Bartolo}}]{Bricard2013}%
  \BibitemOpen
  \bibfield  {author} {\bibinfo {author} {\bibfnamefont {Antoine}\ \bibnamefont
  {Bricard}}, \bibinfo {author} {\bibfnamefont {Jean-Baptiste}\ \bibnamefont
  {Caussin}}, \bibinfo {author} {\bibfnamefont {Nicolas}\ \bibnamefont
  {Desreumaux}}, \bibinfo {author} {\bibfnamefont {Olivier}\ \bibnamefont
  {Dauchot}}, \ and\ \bibinfo {author} {\bibfnamefont {Denis}\ \bibnamefont
  {Bartolo}},\ }\bibfield  {title} {\enquote {\bibinfo {title} {Emergence of
  macroscopic directed motion in populations of motile colloids},}\ }\href
  {https://www.nature.com/nature/journal/v503/n7474/full/nature12673.html}
  {\bibfield  {journal} {\bibinfo  {journal} {Nature}\ }\textbf {\bibinfo
  {volume} {503}} (\bibinfo {year} {2013})}\BibitemShut {NoStop}%
\bibitem [{\citenamefont {Yan}\ \emph {et~al.}(2016)\citenamefont {Yan},
  \citenamefont {Han}, \citenamefont {Zhang}, \citenamefont {Xu}, \citenamefont
  {Luijten},\ and\ \citenamefont {Granick}}]{Yan2016}%
  \BibitemOpen
  \bibfield  {author} {\bibinfo {author} {\bibfnamefont {Jing}\ \bibnamefont
  {Yan}}, \bibinfo {author} {\bibfnamefont {Ming}\ \bibnamefont {Han}},
  \bibinfo {author} {\bibfnamefont {Jie}\ \bibnamefont {Zhang}}, \bibinfo
  {author} {\bibfnamefont {Cong}\ \bibnamefont {Xu}}, \bibinfo {author}
  {\bibfnamefont {Erik}\ \bibnamefont {Luijten}}, \ and\ \bibinfo {author}
  {\bibfnamefont {Steve}\ \bibnamefont {Granick}},\ }\bibfield  {title}
  {\enquote {\bibinfo {title} {Reconfiguring active particles by electrostatic
  imbalance},}\ }\href {\doibase https://doi.org/10.1038/nmat4696} {\bibfield
  {journal} {\bibinfo  {journal} {Nature materials}\ }\textbf {\bibinfo
  {volume} {15}},\ \bibinfo {pages} {1095} (\bibinfo {year}
  {2016})}\BibitemShut {NoStop}%
\bibitem [{\citenamefont {Zhang}\ \emph {et~al.}(2017)\citenamefont {Zhang},
  \citenamefont {Luijten}, \citenamefont {Grzybowski},\ and\ \citenamefont
  {Granick}}]{Granick_Review}%
  \BibitemOpen
  \bibfield  {author} {\bibinfo {author} {\bibfnamefont {Jie}\ \bibnamefont
  {Zhang}}, \bibinfo {author} {\bibfnamefont {Erik}\ \bibnamefont {Luijten}},
  \bibinfo {author} {\bibfnamefont {Bartosz~A.}\ \bibnamefont {Grzybowski}}, \
  and\ \bibinfo {author} {\bibfnamefont {Steve}\ \bibnamefont {Granick}},\
  }\bibfield  {title} {\enquote {\bibinfo {title} {Active colloids with
  collective mobility status and research opportunities},}\ }\href {\doibase
  10.1039/C7CS00461C} {\bibfield  {journal} {\bibinfo  {journal} {Chem. Soc.
  Rev.}\ }\textbf {\bibinfo {volume} {46}},\ \bibinfo {pages} {5551--5569}
  (\bibinfo {year} {2017})}\BibitemShut {NoStop}%
\bibitem [{\citenamefont {Deseigne}\ \emph {et~al.}(2012)\citenamefont
  {Deseigne}, \citenamefont {L{\'e}onard}, \citenamefont {Dauchot},\ and\
  \citenamefont {Chat{\'e}}}]{Deseigne2012}%
  \BibitemOpen
  \bibfield  {author} {\bibinfo {author} {\bibfnamefont {Julien}\ \bibnamefont
  {Deseigne}}, \bibinfo {author} {\bibfnamefont {S{\'e}bastien}\ \bibnamefont
  {L{\'e}onard}}, \bibinfo {author} {\bibfnamefont {Olivier}\ \bibnamefont
  {Dauchot}}, \ and\ \bibinfo {author} {\bibfnamefont {Hugues}\ \bibnamefont
  {Chat{\'e}}},\ }\bibfield  {title} {\enquote {\bibinfo {title} {Vibrated
  polar disks: spontaneous motion, binary collisions, and collective
  dynamics},}\ }\href {\doibase https://doi.org/10.1039/C2SM25186H} {\bibfield
  {journal} {\bibinfo  {journal} {Soft Matter}\ }\textbf {\bibinfo {volume}
  {8}},\ \bibinfo {pages} {5629--5639} (\bibinfo {year} {2012})}\BibitemShut
  {NoStop}%
\bibitem [{\citenamefont {Kumar}\ \emph {et~al.}(2014)\citenamefont {Kumar},
  \citenamefont {Soni}, \citenamefont {Ramaswamy},\ and\ \citenamefont
  {Sood}}]{Sood2014}%
  \BibitemOpen
  \bibfield  {author} {\bibinfo {author} {\bibfnamefont {Nitin}\ \bibnamefont
  {Kumar}}, \bibinfo {author} {\bibfnamefont {Harsh}\ \bibnamefont {Soni}},
  \bibinfo {author} {\bibfnamefont {Sriram}\ \bibnamefont {Ramaswamy}}, \ and\
  \bibinfo {author} {\bibfnamefont {AK}~\bibnamefont {Sood}},\ }\bibfield
  {title} {\enquote {\bibinfo {title} {Flocking at a distance in active
  granular matter},}\ }\href {https://www.nature.com/articles/ncomms5688}
  {\bibfield  {journal} {\bibinfo  {journal} {Nature Communications}\ }\textbf
  {\bibinfo {volume} {5}} (\bibinfo {year} {2014})}\BibitemShut {NoStop}%
\bibitem [{\citenamefont {Doostmohammadi}\ \emph {et~al.}(2018)\citenamefont
  {Doostmohammadi}, \citenamefont {Ign{\'e}s-Mullol}, \citenamefont {Yeomans},\
  and\ \citenamefont {Sagu{\'e}s}}]{Doostmohammadi2018}%
  \BibitemOpen
  \bibfield  {author} {\bibinfo {author} {\bibfnamefont {Amin}\ \bibnamefont
  {Doostmohammadi}}, \bibinfo {author} {\bibfnamefont {Jordi}\ \bibnamefont
  {Ign{\'e}s-Mullol}}, \bibinfo {author} {\bibfnamefont {Julia~M}\ \bibnamefont
  {Yeomans}}, \ and\ \bibinfo {author} {\bibfnamefont {Francesc}\ \bibnamefont
  {Sagu{\'e}s}},\ }\bibfield  {title} {\enquote {\bibinfo {title} {Active
  nematics},}\ }\href {\doibase https://doi.org/10.1038/s41467-018-05666-8}
  {\bibfield  {journal} {\bibinfo  {journal} {Nature communications}\ }\textbf
  {\bibinfo {volume} {9}},\ \bibinfo {pages} {1--13} (\bibinfo {year}
  {2018})}\BibitemShut {NoStop}%
\bibitem [{\citenamefont {Shankar}\ \emph
  {et~al.}(2018{\natexlab{a}})\citenamefont {Shankar}, \citenamefont
  {Ramaswamy},\ and\ \citenamefont {Marchetti}}]{shankar2018low}%
  \BibitemOpen
  \bibfield  {author} {\bibinfo {author} {\bibfnamefont {Suraj}\ \bibnamefont
  {Shankar}}, \bibinfo {author} {\bibfnamefont {Sriram}\ \bibnamefont
  {Ramaswamy}}, \ and\ \bibinfo {author} {\bibfnamefont {M~Cristina}\
  \bibnamefont {Marchetti}},\ }\bibfield  {title} {\enquote {\bibinfo {title}
  {Low-noise phase of a two-dimensional active nematic system},}\ }\href@noop
  {} {\bibfield  {journal} {\bibinfo  {journal} {Physical Review E}\ }\textbf
  {\bibinfo {volume} {97}},\ \bibinfo {pages} {012707} (\bibinfo {year}
  {2018}{\natexlab{a}})}\BibitemShut {NoStop}%
\bibitem [{\citenamefont {Sanchez}\ \emph {et~al.}(2012)\citenamefont
  {Sanchez}, \citenamefont {Chen}, \citenamefont {DeCamp}, \citenamefont
  {Heymann},\ and\ \citenamefont {Dogic}}]{Sanchez2012}%
  \BibitemOpen
  \bibfield  {author} {\bibinfo {author} {\bibfnamefont {Tim}\ \bibnamefont
  {Sanchez}}, \bibinfo {author} {\bibfnamefont {Daniel T.~N.}\ \bibnamefont
  {Chen}}, \bibinfo {author} {\bibfnamefont {Stephen~J.}\ \bibnamefont
  {DeCamp}}, \bibinfo {author} {\bibfnamefont {Michael}\ \bibnamefont
  {Heymann}}, \ and\ \bibinfo {author} {\bibfnamefont {Zvonimir}\ \bibnamefont
  {Dogic}},\ }\bibfield  {title} {\enquote {\bibinfo {title} {Spontaneous
  motion in hierarchically assembled active matter},}\ }\href
  {http://dx.doi.org/10.1038/nature11591} {\bibfield  {journal} {\bibinfo
  {journal} {Nature}\ }\textbf {\bibinfo {volume} {491}},\ \bibinfo {pages}
  {431--434} (\bibinfo {year} {2012})}\BibitemShut {NoStop}%
\bibitem [{\citenamefont {Giomi}\ \emph {et~al.}(2013)\citenamefont {Giomi},
  \citenamefont {Bowick}, \citenamefont {Ma},\ and\ \citenamefont
  {Marchetti}}]{giomi2013defect}%
  \BibitemOpen
  \bibfield  {author} {\bibinfo {author} {\bibfnamefont {Luca}\ \bibnamefont
  {Giomi}}, \bibinfo {author} {\bibfnamefont {Mark~J}\ \bibnamefont {Bowick}},
  \bibinfo {author} {\bibfnamefont {Xu}~\bibnamefont {Ma}}, \ and\ \bibinfo
  {author} {\bibfnamefont {M~Cristina}\ \bibnamefont {Marchetti}},\ }\bibfield
  {title} {\enquote {\bibinfo {title} {Defect annihilation and proliferation in
  active nematics},}\ }\href@noop {} {\bibfield  {journal} {\bibinfo  {journal}
  {Physical review letters}\ }\textbf {\bibinfo {volume} {110}},\ \bibinfo
  {pages} {228101} (\bibinfo {year} {2013})}\BibitemShut {NoStop}%
\bibitem [{\citenamefont {Shankar}\ \emph
  {et~al.}(2018{\natexlab{b}})\citenamefont {Shankar}, \citenamefont
  {Ramaswamy}, \citenamefont {Marchetti},\ and\ \citenamefont
  {Bowick}}]{shankar2018defect}%
  \BibitemOpen
  \bibfield  {author} {\bibinfo {author} {\bibfnamefont {Suraj}\ \bibnamefont
  {Shankar}}, \bibinfo {author} {\bibfnamefont {Sriram}\ \bibnamefont
  {Ramaswamy}}, \bibinfo {author} {\bibfnamefont {M~Cristina}\ \bibnamefont
  {Marchetti}}, \ and\ \bibinfo {author} {\bibfnamefont {Mark~J}\ \bibnamefont
  {Bowick}},\ }\bibfield  {title} {\enquote {\bibinfo {title} {Defect unbinding
  in active nematics},}\ }\href@noop {} {\bibfield  {journal} {\bibinfo
  {journal} {Physical review letters}\ }\textbf {\bibinfo {volume} {121}},\
  \bibinfo {pages} {108002} (\bibinfo {year} {2018}{\natexlab{b}})}\BibitemShut
  {NoStop}%
\bibitem [{\citenamefont {Blackman}\ and\ \citenamefont
  {Tag{\"u}e{\~n}a}(1991)}]{Blackman1991}%
  \BibitemOpen
  \bibfield  {author} {\bibinfo {author} {\bibfnamefont {John~A}\ \bibnamefont
  {Blackman}}\ and\ \bibinfo {author} {\bibfnamefont {J}~\bibnamefont
  {Tag{\"u}e{\~n}a}},\ }\href@noop {} {\emph {\bibinfo {title} {Disorder in
  condensed matter physics: a volume in honour of Roger Elliott}}}\ (\bibinfo
  {publisher} {Oxford University Press, USA},\ \bibinfo {year}
  {1991})\BibitemShut {NoStop}%
\bibitem [{\citenamefont {De~Gennes}(2018)}]{deGennes2018}%
  \BibitemOpen
  \bibfield  {author} {\bibinfo {author} {\bibfnamefont {Pierre-Gilles}\
  \bibnamefont {De~Gennes}},\ }\href@noop {} {\emph {\bibinfo {title}
  {Superconductivity of metals and alloys}}}\ (\bibinfo  {publisher} {CRC
  Press},\ \bibinfo {year} {2018})\BibitemShut {NoStop}%
\bibitem [{\citenamefont {Crabtree}\ and\ \citenamefont
  {Nelson}(1997)}]{Crabtree1997}%
  \BibitemOpen
  \bibfield  {author} {\bibinfo {author} {\bibfnamefont {George~W}\
  \bibnamefont {Crabtree}}\ and\ \bibinfo {author} {\bibfnamefont {David~R}\
  \bibnamefont {Nelson}},\ }\bibfield  {title} {\enquote {\bibinfo {title}
  {Vortex physics in high-temperature superconductors},}\ }\href@noop {}
  {\bibfield  {journal} {\bibinfo  {journal} {Physics Today}\ }\textbf
  {\bibinfo {volume} {50}},\ \bibinfo {pages} {38--45} (\bibinfo {year}
  {1997})}\BibitemShut {NoStop}%
\bibitem [{\citenamefont {Chepizhko}\ \emph {et~al.}(2013)\citenamefont
  {Chepizhko}, \citenamefont {Altmann},\ and\ \citenamefont
  {Peruani}}]{Peruani2013}%
  \BibitemOpen
  \bibfield  {author} {\bibinfo {author} {\bibfnamefont {Oleksandr}\
  \bibnamefont {Chepizhko}}, \bibinfo {author} {\bibfnamefont {Eduardo~G.}\
  \bibnamefont {Altmann}}, \ and\ \bibinfo {author} {\bibfnamefont {Fernando}\
  \bibnamefont {Peruani}},\ }\bibfield  {title} {\enquote {\bibinfo {title}
  {Optimal noise maximizes collective motion in heterogeneous media},}\ }\href
  {\doibase 10.1103/PhysRevLett.110.238101} {\bibfield  {journal} {\bibinfo
  {journal} {Phys. Rev. Lett.}\ }\textbf {\bibinfo {volume} {110}},\ \bibinfo
  {pages} {238101} (\bibinfo {year} {2013})}\BibitemShut {NoStop}%
\bibitem [{\citenamefont {Quint}\ and\ \citenamefont
  {Gopinathan}(2015)}]{Quint2015}%
  \BibitemOpen
  \bibfield  {author} {\bibinfo {author} {\bibfnamefont {David~A}\ \bibnamefont
  {Quint}}\ and\ \bibinfo {author} {\bibfnamefont {Ajay}\ \bibnamefont
  {Gopinathan}},\ }\bibfield  {title} {\enquote {\bibinfo {title}
  {Topologically induced swarming phase transition on a 2d percolated
  lattice},}\ }\href {\doibase https://doi.org/10.1088/1478-3975/12/4/046008}
  {\bibfield  {journal} {\bibinfo  {journal} {Physical biology}\ }\textbf
  {\bibinfo {volume} {12}},\ \bibinfo {pages} {046008} (\bibinfo {year}
  {2015})}\BibitemShut {NoStop}%
\bibitem [{\citenamefont {Bechinger}\ \emph {et~al.}(2016)\citenamefont
  {Bechinger}, \citenamefont {Di~Leonardo}, \citenamefont {L{\"o}wen},
  \citenamefont {Reichhardt}, \citenamefont {Volpe},\ and\ \citenamefont
  {Volpe}}]{Bechinger2016}%
  \BibitemOpen
  \bibfield  {author} {\bibinfo {author} {\bibfnamefont {Clemens}\ \bibnamefont
  {Bechinger}}, \bibinfo {author} {\bibfnamefont {Roberto}\ \bibnamefont
  {Di~Leonardo}}, \bibinfo {author} {\bibfnamefont {Hartmut}\ \bibnamefont
  {L{\"o}wen}}, \bibinfo {author} {\bibfnamefont {Charles}\ \bibnamefont
  {Reichhardt}}, \bibinfo {author} {\bibfnamefont {Giorgio}\ \bibnamefont
  {Volpe}}, \ and\ \bibinfo {author} {\bibfnamefont {Giovanni}\ \bibnamefont
  {Volpe}},\ }\bibfield  {title} {\enquote {\bibinfo {title} {Active particles
  in complex and crowded environments},}\ }\href@noop {} {\bibfield  {journal}
  {\bibinfo  {journal} {Reviews of Modern Physics}\ }\textbf {\bibinfo {volume}
  {88}},\ \bibinfo {pages} {045006} (\bibinfo {year} {2016})}\BibitemShut
  {NoStop}%
\bibitem [{\citenamefont {Morin}\ \emph
  {et~al.}(2017{\natexlab{a}})\citenamefont {Morin}, \citenamefont
  {Desreumaux}, \citenamefont {Caussin},\ and\ \citenamefont
  {Bartolo}}]{Morin2017}%
  \BibitemOpen
  \bibfield  {author} {\bibinfo {author} {\bibfnamefont {Alexandre}\
  \bibnamefont {Morin}}, \bibinfo {author} {\bibfnamefont {Nicolas}\
  \bibnamefont {Desreumaux}}, \bibinfo {author} {\bibfnamefont {Jean-Baptiste}\
  \bibnamefont {Caussin}}, \ and\ \bibinfo {author} {\bibfnamefont {Denis}\
  \bibnamefont {Bartolo}},\ }\bibfield  {title} {\enquote {\bibinfo {title}
  {Distortion and destruction of colloidal flocks in disordered
  environments},}\ }\href {\doibase https://doi.org/10.1038/nphys3903}
  {\bibfield  {journal} {\bibinfo  {journal} {Nature Physics}\ }\textbf
  {\bibinfo {volume} {13}},\ \bibinfo {pages} {63} (\bibinfo {year}
  {2017}{\natexlab{a}})}\BibitemShut {NoStop}%
\bibitem [{\citenamefont {Das}\ \emph {et~al.}(2018)\citenamefont {Das},
  \citenamefont {Kumar},\ and\ \citenamefont {Mishra}}]{Das2018}%
  \BibitemOpen
  \bibfield  {author} {\bibinfo {author} {\bibfnamefont {Rakesh}\ \bibnamefont
  {Das}}, \bibinfo {author} {\bibfnamefont {Manoranjan}\ \bibnamefont {Kumar}},
  \ and\ \bibinfo {author} {\bibfnamefont {Shradha}\ \bibnamefont {Mishra}},\
  }\bibfield  {title} {\enquote {\bibinfo {title} {Polar flock in the presence
  of random quenched rotators},}\ }\href {\doibase 10.1103/PhysRevE.98.060602}
  {\bibfield  {journal} {\bibinfo  {journal} {Phys. Rev. E}\ }\textbf {\bibinfo
  {volume} {98}},\ \bibinfo {pages} {060602} (\bibinfo {year}
  {2018})}\BibitemShut {NoStop}%
\bibitem [{\citenamefont {Reichhardt}\ and\ \citenamefont
  {Reichhardt}(2018)}]{Reichhardt2018}%
  \BibitemOpen
  \bibfield  {author} {\bibinfo {author} {\bibfnamefont {CJ~Olson}\
  \bibnamefont {Reichhardt}}\ and\ \bibinfo {author} {\bibfnamefont {Charles}\
  \bibnamefont {Reichhardt}},\ }\bibfield  {title} {\enquote {\bibinfo {title}
  {Avalanche dynamics for active matter in heterogeneous media},}\ }\href
  {\doibase https://doi.org/10.1088/1367-2630/aaa392} {\bibfield  {journal}
  {\bibinfo  {journal} {New Journal of Physics}\ }\textbf {\bibinfo {volume}
  {20}},\ \bibinfo {pages} {025002} (\bibinfo {year} {2018})}\BibitemShut
  {NoStop}%
\bibitem [{\citenamefont {Toner}\ \emph
  {et~al.}(2018{\natexlab{a}})\citenamefont {Toner}, \citenamefont
  {Guttenberg},\ and\ \citenamefont {Tu}}]{Toner2018}%
  \BibitemOpen
  \bibfield  {author} {\bibinfo {author} {\bibfnamefont {John}\ \bibnamefont
  {Toner}}, \bibinfo {author} {\bibfnamefont {Nicholas}\ \bibnamefont
  {Guttenberg}}, \ and\ \bibinfo {author} {\bibfnamefont {Yuhai}\ \bibnamefont
  {Tu}},\ }\bibfield  {title} {\enquote {\bibinfo {title} {Swarming in the
  dirt: Ordered flocks with quenched disorder},}\ }\href {\doibase
  10.1103/PhysRevLett.121.248002} {\bibfield  {journal} {\bibinfo  {journal}
  {Phys. Rev. Lett.}\ }\textbf {\bibinfo {volume} {121}},\ \bibinfo {pages}
  {248002} (\bibinfo {year} {2018}{\natexlab{a}})}\BibitemShut {NoStop}%
\bibitem [{\citenamefont {Toner}\ \emph
  {et~al.}(2018{\natexlab{b}})\citenamefont {Toner}, \citenamefont
  {Guttenberg},\ and\ \citenamefont {Tu}}]{Toner20182}%
  \BibitemOpen
  \bibfield  {author} {\bibinfo {author} {\bibfnamefont {John}\ \bibnamefont
  {Toner}}, \bibinfo {author} {\bibfnamefont {Nicholas}\ \bibnamefont
  {Guttenberg}}, \ and\ \bibinfo {author} {\bibfnamefont {Yuhai}\ \bibnamefont
  {Tu}},\ }\bibfield  {title} {\enquote {\bibinfo {title} {Hydrodynamic theory
  of flocking in the presence of quenched disorder},}\ }\href {\doibase
  10.1103/PhysRevE.98.062604} {\bibfield  {journal} {\bibinfo  {journal} {Phys.
  Rev. E}\ }\textbf {\bibinfo {volume} {98}},\ \bibinfo {pages} {062604}
  (\bibinfo {year} {2018}{\natexlab{b}})}\BibitemShut {NoStop}%
\bibitem [{\citenamefont {Rubinstein}\ \emph {et~al.}(1983)\citenamefont
  {Rubinstein}, \citenamefont {Shraiman},\ and\ \citenamefont
  {Nelson}}]{Rubinstein83}%
  \BibitemOpen
  \bibfield  {author} {\bibinfo {author} {\bibfnamefont {Michael}\ \bibnamefont
  {Rubinstein}}, \bibinfo {author} {\bibfnamefont {Boris}\ \bibnamefont
  {Shraiman}}, \ and\ \bibinfo {author} {\bibfnamefont {David~R}\ \bibnamefont
  {Nelson}},\ }\bibfield  {title} {\enquote {\bibinfo {title} {Two-dimensional
  xy magnets with random dzyaloshinskii-moriya interactions},}\ }\href
  {\doibase 10.1103/PhysRevB.27.1800} {\bibfield  {journal} {\bibinfo
  {journal} {Physical Review B}\ }\textbf {\bibinfo {volume} {27}},\ \bibinfo
  {pages} {1800} (\bibinfo {year} {1983})}\BibitemShut {NoStop}%
\bibitem [{\citenamefont {Nattermann}\ and\ \citenamefont
  {Scheidl}(2000)}]{Nattermann2000}%
  \BibitemOpen
  \bibfield  {author} {\bibinfo {author} {\bibfnamefont {Thomas}\ \bibnamefont
  {Nattermann}}\ and\ \bibinfo {author} {\bibfnamefont {Stefan}\ \bibnamefont
  {Scheidl}},\ }\bibfield  {title} {\enquote {\bibinfo {title} {Vortex-glass
  phases in type-ii superconductors},}\ }\href {\doibase
  10.1080/000187300412257} {\bibfield  {journal} {\bibinfo  {journal} {Advances
  in Physics}\ }\textbf {\bibinfo {volume} {49}},\ \bibinfo {pages} {607--704}
  (\bibinfo {year} {2000})}\BibitemShut {NoStop}%
\bibitem [{\citenamefont {Carpentier}\ and\ \citenamefont
  {Le~Doussal}(2000)}]{Carpentier2000}%
  \BibitemOpen
  \bibfield  {author} {\bibinfo {author} {\bibfnamefont {David}\ \bibnamefont
  {Carpentier}}\ and\ \bibinfo {author} {\bibfnamefont {Pierre}\ \bibnamefont
  {Le~Doussal}},\ }\bibfield  {title} {\enquote {\bibinfo {title} {Topological
  transitions and freezing in xy models and coulomb gases with quenched
  disorder: renormalization via traveling waves},}\ }\href {\doibase
  10.1016/S0550-3213(00)00468-5} {\bibfield  {journal} {\bibinfo  {journal}
  {Nuclear Physics B}\ }\textbf {\bibinfo {volume} {588}},\ \bibinfo {pages}
  {565--629} (\bibinfo {year} {2000})}\BibitemShut {NoStop}%
\bibitem [{\citenamefont {Quincke}(1896)}]{Quincke}%
  \BibitemOpen
  \bibfield  {author} {\bibinfo {author} {\bibfnamefont {G.}~\bibnamefont
  {Quincke}},\ }\bibfield  {title} {\enquote {\bibinfo {title} {Uber rotationen
  im constanten electrischen felde},}\ }\href
  {http://onlinelibrary.wiley.com/doi/10.1002/andp.18962951102/abstract}
  {\bibfield  {journal} {\bibinfo  {journal} {Annalen der Physik}\ ,\ \bibinfo
  {pages} {417--486}} (\bibinfo {year} {1896})}\BibitemShut {NoStop}%
\bibitem [{\citenamefont {Lavrentovich}(2016)}]{Lavrentovich2016}%
  \BibitemOpen
  \bibfield  {author} {\bibinfo {author} {\bibfnamefont {Oleg~D}\ \bibnamefont
  {Lavrentovich}},\ }\bibfield  {title} {\enquote {\bibinfo {title} {Active
  colloids in liquid crystals},}\ }\href@noop {} {\bibfield  {journal}
  {\bibinfo  {journal} {Current opinion in colloid \& interface science}\
  }\textbf {\bibinfo {volume} {21}},\ \bibinfo {pages} {97--109} (\bibinfo
  {year} {2016})}\BibitemShut {NoStop}%
\bibitem [{\citenamefont {Bricard}\ \emph {et~al.}(2015)\citenamefont
  {Bricard}, \citenamefont {Caussin}, \citenamefont {Das}, \citenamefont
  {Savoie}, \citenamefont {Chikkadi}, \citenamefont {Shitara}, \citenamefont
  {Chepizhko}, \citenamefont {Peruani}, \citenamefont {Saintillan},\ and\
  \citenamefont {Bartolo}}]{Bricard2015}%
  \BibitemOpen
  \bibfield  {author} {\bibinfo {author} {\bibfnamefont {Antoine}\ \bibnamefont
  {Bricard}}, \bibinfo {author} {\bibfnamefont {Jean-Baptiste}\ \bibnamefont
  {Caussin}}, \bibinfo {author} {\bibfnamefont {Debasish}\ \bibnamefont {Das}},
  \bibinfo {author} {\bibfnamefont {Charles}\ \bibnamefont {Savoie}}, \bibinfo
  {author} {\bibfnamefont {Vijayakumar}\ \bibnamefont {Chikkadi}}, \bibinfo
  {author} {\bibfnamefont {Kyohei}\ \bibnamefont {Shitara}}, \bibinfo {author}
  {\bibfnamefont {Oleksandr}\ \bibnamefont {Chepizhko}}, \bibinfo {author}
  {\bibfnamefont {Fernando}\ \bibnamefont {Peruani}}, \bibinfo {author}
  {\bibfnamefont {David}\ \bibnamefont {Saintillan}}, \ and\ \bibinfo {author}
  {\bibfnamefont {Denis}\ \bibnamefont {Bartolo}},\ }\bibfield  {title}
  {\enquote {\bibinfo {title} {Emergent vortices in populations of colloidal
  rollers},}\ }\href {\doibase 10.1038/ncomms8470} {\bibfield  {journal}
  {\bibinfo  {journal} {Nature communications}\ }\textbf {\bibinfo {volume}
  {6}},\ \bibinfo {pages} {7470} (\bibinfo {year} {2015})}\BibitemShut
  {NoStop}%
\bibitem [{\citenamefont {Geyer}\ \emph {et~al.}(2019)\citenamefont {Geyer},
  \citenamefont {Martin}, \citenamefont {Tailleur},\ and\ \citenamefont
  {Bartolo}}]{Geyer2019}%
  \BibitemOpen
  \bibfield  {author} {\bibinfo {author} {\bibfnamefont {Delphine}\
  \bibnamefont {Geyer}}, \bibinfo {author} {\bibfnamefont {David}\ \bibnamefont
  {Martin}}, \bibinfo {author} {\bibfnamefont {Julien}\ \bibnamefont
  {Tailleur}}, \ and\ \bibinfo {author} {\bibfnamefont {Denis}\ \bibnamefont
  {Bartolo}},\ }\bibfield  {title} {\enquote {\bibinfo {title} {Freezing a
  flock: Motility-induced phase separation in polar active liquids},}\ }\href
  {https://doi.org/10.1103/PhysRevX.9.031043} {\bibfield  {journal} {\bibinfo
  {journal} {Physical Review X}\ }\textbf {\bibinfo {volume} {9}},\ \bibinfo
  {pages} {031043} (\bibinfo {year} {2019})}\BibitemShut {NoStop}%
\bibitem [{\citenamefont {Toner}\ and\ \citenamefont {Tu}(1995)}]{Toner95}%
  \BibitemOpen
  \bibfield  {author} {\bibinfo {author} {\bibfnamefont {John}\ \bibnamefont
  {Toner}}\ and\ \bibinfo {author} {\bibfnamefont {Yuhai}\ \bibnamefont {Tu}},\
  }\bibfield  {title} {\enquote {\bibinfo {title} {Long-range order in a
  two-dimensional dynamical $\mathrm{XY}$ model: How birds fly together},}\
  }\href {\doibase 10.1103/PhysRevLett.75.4326} {\bibfield  {journal} {\bibinfo
   {journal} {Phys. Rev. Lett.}\ }\textbf {\bibinfo {volume} {75}},\ \bibinfo
  {pages} {4326--4329} (\bibinfo {year} {1995})}\BibitemShut {NoStop}%
\bibitem [{\citenamefont {Zeitz}\ \emph {et~al.}(2017)\citenamefont {Zeitz},
  \citenamefont {Wolff},\ and\ \citenamefont {Stark}}]{Zeitz2017}%
  \BibitemOpen
  \bibfield  {author} {\bibinfo {author} {\bibfnamefont {Maria}\ \bibnamefont
  {Zeitz}}, \bibinfo {author} {\bibfnamefont {Katrin}\ \bibnamefont {Wolff}}, \
  and\ \bibinfo {author} {\bibfnamefont {Holger}\ \bibnamefont {Stark}},\
  }\bibfield  {title} {\enquote {\bibinfo {title} {Active brownian particles
  moving in a random lorentz gas},}\ }\href@noop {} {\bibfield  {journal}
  {\bibinfo  {journal} {The European Physical Journal E}\ }\textbf {\bibinfo
  {volume} {40}},\ \bibinfo {pages} {23} (\bibinfo {year} {2017})}\BibitemShut
  {NoStop}%
\bibitem [{\citenamefont {Morin}\ \emph
  {et~al.}(2017{\natexlab{b}})\citenamefont {Morin}, \citenamefont
  {Lopes~Cardozo}, \citenamefont {Chikkadi},\ and\ \citenamefont
  {Bartolo}}]{Morin2017b}%
  \BibitemOpen
  \bibfield  {author} {\bibinfo {author} {\bibfnamefont {Alexandre}\
  \bibnamefont {Morin}}, \bibinfo {author} {\bibfnamefont {David}\ \bibnamefont
  {Lopes~Cardozo}}, \bibinfo {author} {\bibfnamefont {Vijayakumar}\
  \bibnamefont {Chikkadi}}, \ and\ \bibinfo {author} {\bibfnamefont {Denis}\
  \bibnamefont {Bartolo}},\ }\bibfield  {title} {\enquote {\bibinfo {title}
  {Diffusion, subdiffusion, and localization of active colloids in random post
  lattices},}\ }\href {\doibase 10.1103/PhysRevE.96.042611} {\bibfield
  {journal} {\bibinfo  {journal} {Phys. Rev. E}\ }\textbf {\bibinfo {volume}
  {96}},\ \bibinfo {pages} {042611} (\bibinfo {year}
  {2017}{\natexlab{b}})}\BibitemShut {NoStop}%
\bibitem [{\citenamefont {Wioland}\ \emph {et~al.}(2016)\citenamefont
  {Wioland}, \citenamefont {Lushi},\ and\ \citenamefont
  {Goldstein}}]{Wioland2016}%
  \BibitemOpen
  \bibfield  {author} {\bibinfo {author} {\bibfnamefont {Hugo}\ \bibnamefont
  {Wioland}}, \bibinfo {author} {\bibfnamefont {Enkeleida}\ \bibnamefont
  {Lushi}}, \ and\ \bibinfo {author} {\bibfnamefont {Raymond~E}\ \bibnamefont
  {Goldstein}},\ }\bibfield  {title} {\enquote {\bibinfo {title} {Directed
  collective motion of bacteria under channel confinement},}\ }\href {\doibase
  https://doi.org/10.1088/1367-2630/18/7/075002} {\bibfield  {journal}
  {\bibinfo  {journal} {New Journal of Physics}\ }\textbf {\bibinfo {volume}
  {18}},\ \bibinfo {pages} {075002} (\bibinfo {year} {2016})}\BibitemShut
  {NoStop}%
\bibitem [{\citenamefont {Wu}\ \emph {et~al.}(2017)\citenamefont {Wu},
  \citenamefont {Hishamunda}, \citenamefont {Chen}, \citenamefont {DeCamp},
  \citenamefont {Chang}, \citenamefont {Fern{\'a}ndez-Nieves}, \citenamefont
  {Fraden},\ and\ \citenamefont {Dogic}}]{Wu2017}%
  \BibitemOpen
  \bibfield  {author} {\bibinfo {author} {\bibfnamefont {Kun-Ta}\ \bibnamefont
  {Wu}}, \bibinfo {author} {\bibfnamefont {Jean~Bernard}\ \bibnamefont
  {Hishamunda}}, \bibinfo {author} {\bibfnamefont {Daniel~TN}\ \bibnamefont
  {Chen}}, \bibinfo {author} {\bibfnamefont {Stephen~J}\ \bibnamefont
  {DeCamp}}, \bibinfo {author} {\bibfnamefont {Ya-Wen}\ \bibnamefont {Chang}},
  \bibinfo {author} {\bibfnamefont {Alberto}\ \bibnamefont
  {Fern{\'a}ndez-Nieves}}, \bibinfo {author} {\bibfnamefont {Seth}\
  \bibnamefont {Fraden}}, \ and\ \bibinfo {author} {\bibfnamefont {Zvonimir}\
  \bibnamefont {Dogic}},\ }\bibfield  {title} {\enquote {\bibinfo {title}
  {Transition from turbulent to coherent flows in confined three-dimensional
  active fluids},}\ }\href {\doibase 10.1126/science.aal1979} {\bibfield
  {journal} {\bibinfo  {journal} {Science}\ }\textbf {\bibinfo {volume}
  {355}},\ \bibinfo {pages} {eaal1979} (\bibinfo {year} {2017})}\BibitemShut
  {NoStop}%
\bibitem [{\citenamefont {Opathalage}\ \emph {et~al.}(2019)\citenamefont
  {Opathalage}, \citenamefont {Norton}, \citenamefont {Juniper}, \citenamefont
  {Langeslay}, \citenamefont {Aghvami}, \citenamefont {Fraden},\ and\
  \citenamefont {Dogic}}]{Opathalage2019}%
  \BibitemOpen
  \bibfield  {author} {\bibinfo {author} {\bibfnamefont {Achini}\ \bibnamefont
  {Opathalage}}, \bibinfo {author} {\bibfnamefont {Michael~M}\ \bibnamefont
  {Norton}}, \bibinfo {author} {\bibfnamefont {Michael~PN}\ \bibnamefont
  {Juniper}}, \bibinfo {author} {\bibfnamefont {Blake}\ \bibnamefont
  {Langeslay}}, \bibinfo {author} {\bibfnamefont {S~Ali}\ \bibnamefont
  {Aghvami}}, \bibinfo {author} {\bibfnamefont {Seth}\ \bibnamefont {Fraden}},
  \ and\ \bibinfo {author} {\bibfnamefont {Zvonimir}\ \bibnamefont {Dogic}},\
  }\bibfield  {title} {\enquote {\bibinfo {title} {Self-organized dynamics and
  the transition to turbulence of confined active nematics},}\ }\href {\doibase
  10.1073/pnas.1816733116} {\bibfield  {journal} {\bibinfo  {journal}
  {Proceedings of the National Academy of Sciences}\ }\textbf {\bibinfo
  {volume} {116}},\ \bibinfo {pages} {4788--4797} (\bibinfo {year}
  {2019})}\BibitemShut {NoStop}%
\bibitem [{\citenamefont {Cavagna}(2009)}]{Cavagna2009}%
  \BibitemOpen
  \bibfield  {author} {\bibinfo {author} {\bibfnamefont {Andrea}\ \bibnamefont
  {Cavagna}},\ }\bibfield  {title} {\enquote {\bibinfo {title} {Supercooled
  liquids for pedestrians},}\ }\href {\doibase
  https://doi.org/10.1016/j.physrep.2009.03.003} {\bibfield  {journal}
  {\bibinfo  {journal} {Physics Reports}\ }\textbf {\bibinfo {volume} {476}},\
  \bibinfo {pages} {51--124} (\bibinfo {year} {2009})}\BibitemShut {NoStop}%
\bibitem [{\citenamefont {Fisher}(1989)}]{Fisher89}%
  \BibitemOpen
  \bibfield  {author} {\bibinfo {author} {\bibfnamefont {Matthew P.~A.}\
  \bibnamefont {Fisher}},\ }\bibfield  {title} {\enquote {\bibinfo {title}
  {Vortex-glass superconductivity: A possible new phase in bulk
  high-${\mathrm{t}}_{\mathrm{c}}$ oxides},}\ }\href {\doibase
  10.1103/PhysRevLett.62.1415} {\bibfield  {journal} {\bibinfo  {journal}
  {Phys. Rev. Lett.}\ }\textbf {\bibinfo {volume} {62}},\ \bibinfo {pages}
  {1415--1418} (\bibinfo {year} {1989})}\BibitemShut {NoStop}%
\bibitem [{\citenamefont {Carpentier}\ and\ \citenamefont
  {Le~Doussal}(1998)}]{carpentier1998disordered}%
  \BibitemOpen
  \bibfield  {author} {\bibinfo {author} {\bibfnamefont {David}\ \bibnamefont
  {Carpentier}}\ and\ \bibinfo {author} {\bibfnamefont {Pierre}\ \bibnamefont
  {Le~Doussal}},\ }\bibfield  {title} {\enquote {\bibinfo {title} {Disordered
  xy models and coulomb gases: renormalization via traveling waves},}\ }\href
  {\doibase 10.1103/PhysRevLett.81.2558} {\bibfield  {journal} {\bibinfo
  {journal} {Physical review letters}\ }\textbf {\bibinfo {volume} {81}},\
  \bibinfo {pages} {2558} (\bibinfo {year} {1998})}\BibitemShut {NoStop}%
\bibitem [{\citenamefont {Duclos}\ \emph {et~al.}(2018)\citenamefont {Duclos},
  \citenamefont {Blanch-Mercader}, \citenamefont {Yashunsky}, \citenamefont
  {Salbreux}, \citenamefont {Joanny}, \citenamefont {Prost},\ and\
  \citenamefont {Silberzan}}]{Duclos2018}%
  \BibitemOpen
  \bibfield  {author} {\bibinfo {author} {\bibfnamefont {G}~\bibnamefont
  {Duclos}}, \bibinfo {author} {\bibfnamefont {C}~\bibnamefont
  {Blanch-Mercader}}, \bibinfo {author} {\bibfnamefont {V}~\bibnamefont
  {Yashunsky}}, \bibinfo {author} {\bibfnamefont {G}~\bibnamefont {Salbreux}},
  \bibinfo {author} {\bibfnamefont {J-F}\ \bibnamefont {Joanny}}, \bibinfo
  {author} {\bibfnamefont {J}~\bibnamefont {Prost}}, \ and\ \bibinfo {author}
  {\bibfnamefont {Pascal}\ \bibnamefont {Silberzan}},\ }\bibfield  {title}
  {\enquote {\bibinfo {title} {Spontaneous shear flow in confined cellular
  nematics},}\ }\href {\doibase https://doi.org/10.1038/s41567-018-0099-7}
  {\bibfield  {journal} {\bibinfo  {journal} {Nature physics}\ }\textbf
  {\bibinfo {volume} {14}},\ \bibinfo {pages} {728--732} (\bibinfo {year}
  {2018})}\BibitemShut {NoStop}%
\bibitem [{\citenamefont {Geyer}\ \emph {et~al.}(2018)\citenamefont {Geyer},
  \citenamefont {Morin},\ and\ \citenamefont {Bartolo}}]{Geyer2018}%
  \BibitemOpen
  \bibfield  {author} {\bibinfo {author} {\bibfnamefont {Delphine}\
  \bibnamefont {Geyer}}, \bibinfo {author} {\bibfnamefont {Alexandre}\
  \bibnamefont {Morin}}, \ and\ \bibinfo {author} {\bibfnamefont {Denis}\
  \bibnamefont {Bartolo}},\ }\bibfield  {title} {\enquote {\bibinfo {title}
  {Sounds and hydrodynamics of polar active fluids},}\ }\href
  {https://doi.org/10.1038/s41563-018-0123-4} {\bibfield  {journal} {\bibinfo
  {journal} {Nature materials}\ }\textbf {\bibinfo {volume} {17}},\ \bibinfo
  {pages} {789--793} (\bibinfo {year} {2018})}\BibitemShut {NoStop}%
\bibitem [{\citenamefont {Melcher}\ and\ \citenamefont
  {Taylor}(1969)}]{Taylor69}%
  \BibitemOpen
  \bibfield  {author} {\bibinfo {author} {\bibfnamefont {J.~R.}\ \bibnamefont
  {Melcher}}\ and\ \bibinfo {author} {\bibfnamefont {G.~I.}\ \bibnamefont
  {Taylor}},\ }\bibfield  {title} {\enquote {\bibinfo {title}
  {Electrohydrodynamics: A review of the role of interfacial shear stresses},}\
  }\href {\doibase 10.1146/annurev.fl.01.010169.000551} {\bibfield  {journal}
  {\bibinfo  {journal} {Annual Review of Fluid Mechanics}\ }\textbf {\bibinfo
  {volume} {1}},\ \bibinfo {pages} {111--146} (\bibinfo {year}
  {1969})}\BibitemShut {NoStop}%
\bibitem [{\citenamefont {Lu}\ \emph {et~al.}(2007)\citenamefont {Lu},
  \citenamefont {Sims}, \citenamefont {Oki}, \citenamefont {Macarthur},\ and\
  \citenamefont {Weitz}}]{Lu2007}%
  \BibitemOpen
  \bibfield  {author} {\bibinfo {author} {\bibfnamefont {Peter~J}\ \bibnamefont
  {Lu}}, \bibinfo {author} {\bibfnamefont {Peter~A}\ \bibnamefont {Sims}},
  \bibinfo {author} {\bibfnamefont {Hidekazu}\ \bibnamefont {Oki}}, \bibinfo
  {author} {\bibfnamefont {James~B}\ \bibnamefont {Macarthur}}, \ and\ \bibinfo
  {author} {\bibfnamefont {David~A}\ \bibnamefont {Weitz}},\ }\bibfield
  {title} {\enquote {\bibinfo {title} {Target-locking acquisition with
  real-time confocal (tarc) microscopy},}\ }\href {\doibase
  https://doi.org/10.1364/OE.15.008702} {\bibfield  {journal} {\bibinfo
  {journal} {Optics express}\ }\textbf {\bibinfo {volume} {15}},\ \bibinfo
  {pages} {8702--8712} (\bibinfo {year} {2007})}\BibitemShut {NoStop}%
\bibitem [{\citenamefont {Crocker}\ and\ \citenamefont {Grier}(1996)}]{Grier}%
  \BibitemOpen
  \bibfield  {author} {\bibinfo {author} {\bibfnamefont {John~C}\ \bibnamefont
  {Crocker}}\ and\ \bibinfo {author} {\bibfnamefont {David~G}\ \bibnamefont
  {Grier}},\ }\bibfield  {title} {\enquote {\bibinfo {title} {Methods of
  digital video microscopy for colloidal studies},}\ }\href
  {http://www.sciencedirect.com/science/article/pii/S0021979796902179}
  {\bibfield  {journal} {\bibinfo  {journal} {Journal of colloid and interface
  science}\ }\textbf {\bibinfo {volume} {179}},\ \bibinfo {pages} {298--310}
  (\bibinfo {year} {1996})}\BibitemShut {NoStop}%
\bibitem [{\citenamefont {Blair}\ and\ \citenamefont {Dufresne}()}]{Blair}%
  \BibitemOpen
  \bibfield  {author} {\bibinfo {author} {\bibfnamefont {D.}~\bibnamefont
  {Blair}}\ and\ \bibinfo {author} {\bibfnamefont {E.}~\bibnamefont
  {Dufresne}},\ }\href {http://physics.georgetown.edu/matlab/} {\enquote
  {\bibinfo {title} {The matlab particle tracking code repository},}\
  }\BibitemShut {NoStop}%
\end{thebibliography}

\section*{Methods}
\label{Method}
\subsection{Quincke rollers experiments} 
The experimental setup is similar to that described in~\cite{Geyer2018}.
We disperse polystyrene colloids of radius $a=2.4\,\rm\mu m$ (Thermo Scientific G0500) in a solution of hexadecane including  $5.5\times10^{-2}\,\rm wt\,\%$ of dioctyl sulfosuccinate sodium salt (AOT). We then inject the solution in microfluidic chambers made of two electrodes spaced by a $25\,\rm \mu m$-thick scotch tape. The electrodes are glass slides, coated with indium tin oxide (Solems, ITOSOL30, thickness: $80\,\rm nm$). We let the colloids sediment on the bottom electrode and apply a DC voltage of $120\,\rm V$. The resulting electric field triggers the so-called Quincke electro-rotation and causes the colloids to roll at a constant speed $v_{0}=0.8\,\rm mm/s$~\cite{Bricard2013,Taylor69}.
The microfluidic device is sketched in .~\ref{Fig1}b. We confine the colloidal rollers inside circular chambers of radius $R=1.5\,\rm mm$ including random lattices of circular posts of radius $10\,\rm\mu m$. Both the obstacles and the confining disks are made of a $2\,\rm \mu m$-thick layer of insulating photoresist resin (Microposit S1818) patterned by means of conventional UV-Lithography as explained in~\cite{Morin2017}. The patterns are lithographed on the bottom electrode. Note that the distribution of the obstacle centers corresponds to a planar Poisson process, the circular posts can therefore overlap as see in Fig.~\ref{Fig1}b and Movie S3. The experiments reported in the main text correspond to thirty different microfluidic chambers including obstacle fractions ranging from $0\,\%$ to $38\,\%$. We fine tune the mean roller fraction $\rho_0$ using a T-junction to  inject sequentially a colloidal suspension or a colloid-free solvent.

 If not specified otherwise, the stationary flows are measured in the first chamber $120\,\rm s$ after the application of the DC field. This waiting time is more than one order of magnitude larger that the flows' relaxation time, see Supplementary Note I. For the replica experiments we proceed as follows: the chambers are filled with the colloids at the desired area fraction. We  motorize the colloids, wait for $120\, \rm s$ and film their motion for $5\,\rm s$. We then switch the voltage off, and switch it on again repeating the same procedure fifty times in a row.

\subsection{From Lagrangian trajectories to Eulerian flow fields} 
In order to track of the trajectories of the rollers, we image them with a Nikon AZ100 microscope with a 4.8X magnification and record videos with a CMOS camera (Basler Ace) at $190\,\rm fps$. %
All measurements are systematically repeated three times for different initial conditions and same disorder configuration. If not specified otherwise, we measure all quantities reported in the main text after the ensemble of rollers has reached its stationary state.
\subsubsection{Lagrangian trajectories}
We detect the position of all the rollers with a sub-pixel accuracy using the algorithm introduced by Lu et al in~\cite{Lu2007}. We then reconstruct their trajectories over the whole 3\,mm wide circular chambers using the Crocker and Grier algorithm~\cite{Grier} with the MATLAB routine available at~\cite{Blair}.  %
 We define the individual roller velocities from their displacements over two subsequent frames (time interval: $\delta t=5.3\,\rm ms$):
$\mathbf{v}_{i}(t) = \mathbf{r}_{i}(t+\delta t)-\mathbf{r}_{i}(t)$, where $\mathbf{r}_{i}(t)$ and $\mathbf{v}_{i}(t)$ are respectively the position and velocity of particule $i$ at time $t$. The accuracy of the position measurements is of the order of $0.1\,\rm \mu m$, inducing an accuracy of the order of $40\,\rm \mu m/s$ for individual speed measurements. When powered with an electric field $\mathbf E$ of magnitude $120\,\rm V$, all colloids roll at a constant speed:
\begin{equation}
v_{0}= 0.80 \pm 0.04\,\rm mm/s.
\end{equation} 
In addition, when isolated, their direction of motion freely diffuses on the unit circle with a rotational diffusivity $D_{\rm R}$  defined as the exponential decorrelation rate of the velocity orientation in an isotropic phase:
\begin{equation}
D_{\rm R} = 2.2 \pm 0.1\,\rm  s^{-1}.
\end{equation}

\subsubsection{Eulerian fields}
 Building on these Lagrangian measurements, we reconstruct the instantaneous Eulerian velocity fields $\mathbf{v}(\mathbf{r},t)$ as follows. We average the instantaneous roller velocities in $76.4\,\rm \mu m \times 76.4\,\rm \mu m$ binning windows arranged on a square lattice with a lattice spacing of $15.3\,\rm \mu m$. Given the roller density, each PIV window typically averages the velocity of 25 rollers. We systematically checked that none of our results crucially depends on the specific choice of the size of the binning windows. The polarization and overlap fields are computed with the same spatial resolution from $\mathbf{v}(\mathbf{r},t)$.

To compute the polarization order parameter from the instantaneous velocity field, we first average the radial and azimuthal components of the polarization field $ \bm{p}\rm(\mathbf{r}) \equiv \langle \rm \mathbf{\hat{v}}(\mathbf{r},t) \rangle_{t}$ over square boxes of size $\ell$. We then compute the spatial average: $\bm{p}_{B}\equiv\left(\langle  p_r (\mathbf r)\rangle_{\mathbf r},\langle p_\theta (\mathbf r)\rangle_{\mathbf r}\right)$, in each box $B$. Finally, $\mathcal P(\ell)$ corresponds to the norm of $\bm{p}_{B}$ averaged over all boxes $B$.\\
%%%%%

% Bibliography
%\bibliography{Biblio}
%merlin.mbs apsrev4-1.bst 2010-07-25 4.21a (PWD, AO, DPC) hacked
%Control: key (0)
%Control: author (0) dotless jnrlst
%Control: editor formatted (1) identically to author
%Control: production of article title (0) allowed
%Control: page (1) range
%Control: year (0) verbatim
%Control: production of eprint (0) enabled
%

\end{document}